\documentclass{article}
\usepackage{PRIMEarxiv}
\usepackage[utf8]{inputenc}
\usepackage[T1]{fontenc}
\usepackage{hyperref}
\usepackage{url}
\usepackage{booktabs}
\usepackage{amsfonts}
\usepackage{nicefrac}
\usepackage{microtype}
\usepackage{lipsum}
\usepackage{fancyhdr}
\usepackage{graphicx}
\graphicspath{{media/}}
\usepackage{wrapfig}
\usepackage{tcolorbox}
\usepackage{amsmath}
\usepackage{amssymb}
\usepackage{bbm}
\usepackage{longtable}
\usepackage{array}
\usepackage{subcaption}
\usepackage{multicol}

\title{InterPLM: Discovering Interpretable Features in Protein Language Models via Sparse Autoencoders}

\author{
  Elana Simon\\
  Stanford University\\
  \texttt{epsimon@stanford.edu}
  \And
  James Zou\\
  Stanford University\\
  \texttt{jamesz@stanford.edu}
}

\begin{document}
\maketitle

\begin{abstract}
Protein language models (PLMs) have demonstrated remarkable success in protein modeling and design, yet their internal mechanisms for predicting structure and function remain poorly understood. Here we present a systematic approach to extract and analyze interpretable features from PLMs using sparse autoencoders (SAEs). By training SAEs on embeddings from the PLM ESM-2, we identify up to 2,548 human-interpretable latent features per layer that strongly correlate with up to 143 known biological concepts such as binding sites, structural motifs, and functional domains. In contrast, examining individual neurons in ESM-2 reveals up to 46 neurons per layer with clear conceptual alignment across 15 known concepts, suggesting that PLMs represent most concepts in superposition. Beyond capturing known annotations, we show that ESM-2 learns coherent concepts that do not map onto existing annotations and propose a pipeline using language models to automatically interpret novel latent features learned by the SAEs. As practical applications, we demonstrate how these latent features can fill in missing annotations in protein databases and enable targeted steering of protein sequence generation. Our results demonstrate that PLMs encode rich, interpretable representations of protein biology and we propose a systematic framework to extract and analyze these latent features. In the process, we recover both known biology and potentially new protein motifs. As community resources, we introduce InterPLM (interPLM.ai), an interactive visualization platform for exploring and analyzing learned PLM features, and release code for training and analysis at github.com/ElanaPearl/interPLM.
\end{abstract}


\section{Introduction}
Language models of protein sequences have revolutionized our ability to predict protein structure and function \cite{lin_evolutionary-scale_2023,nijkamp_progen2_2023}. These models achieve their impressive performance by learning rich representations from the vast diversity of naturally evolved proteins. Their success raises a fundamental question: what exactly do protein language models (PLMs) learn? Advances in interpretability research methods now enable us to investigate this question and understand at least some of the representations they have learned in service of this challenging task. 

\begin{figure}[htbp]
    \vspace{-10pt}
    \centering
    \includegraphics[width=\linewidth]{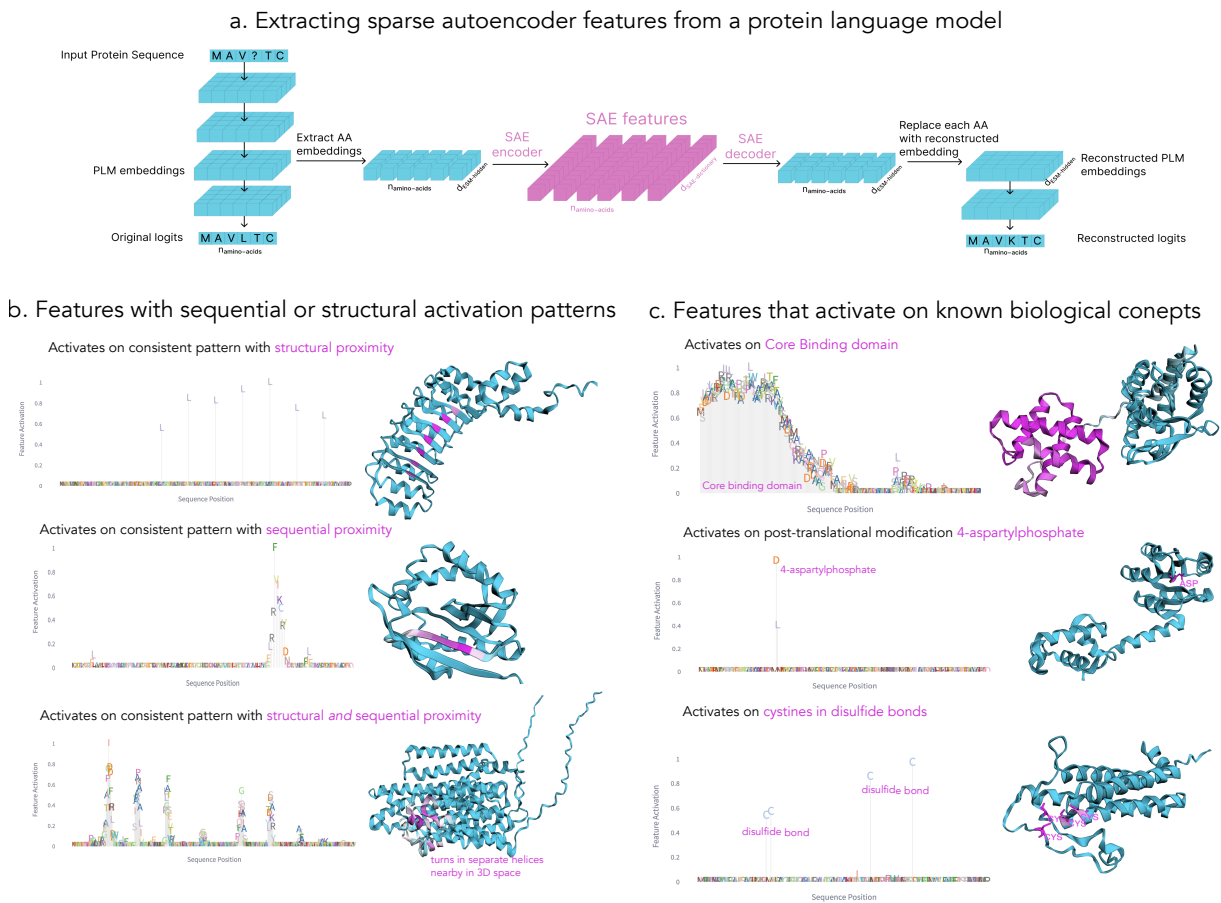}
    \caption{\textbf{Overview of SAE methodology and representative SAE features revealed through automated activation pattern analysis.} a) Pipeline illustrating the extraction of embeddings, their conversion to features, and subsequent reinsertion of reconstructed embeddings into the PLM. b-c) Examples of features exhibiting interpretable activation patterns, both structural and conceptual. Each feature is visualized using a protein where maximal activation occurs. Feature activation intensities are displayed both along the protein sequence (amino acid height indicates activation magnitude) and protein structure (darker pink indicates stronger activation).
    b) Features selected to demonstrate activation patterns in structurally proximate amino acids, sequentially proximate amino acids, or both. c) Features selected based on significant associations between activation patterns and known Swiss-Prot biological concept annotations. Feature identifiers and corresponding layers (top to bottom): Left panel (f/3147, f/10091, f/67, layer 4); Right panel (f/7125, f/8128, f/1455 from layers 4, 5, 5).}
    \label{fig:example_features}
\end{figure}
Gaining insights into the internal mechanisms of PLMs is crucial for model development and biological discovery \cite{zhang_protein_2024}. Understanding how these models process information enables more informed decisions about model design, beyond merely optimizing performance metrics. By analyzing the features driving predictions, we can identify spurious correlations or biases, assess the generalizability of learned representations, and potentially uncover novel biological insights. For instance, models might learn to predict certain protein properties based on subtle patterns or principles that have eluded human analysis, opening up new avenues for experimental investigation. Conversely, detecting biologically implausible features can guide improvements to the models' inductive biases and learning algorithms. This systematic analysis of prediction mechanisms offers opportunities to enhance model performance and reliability while extracting biological hypotheses from learned representations. Furthermore, PLM interpretation can illuminate how much these models learn genuine physical and chemical principles governing protein structure, as opposed to simply memorizing structural motifs. As PLMs continue to advance, systematic interpretation frameworks may uncover increasingly sophisticated biological insights.

In this work, we create a versatile framework using sparse auto-encoders (SAEs) to interpret latent features learned by PLMs. To demonstrate this framework, we analyze amino acid embeddings across all layers of ESM-2 \cite{lin_evolutionary-scale_2023}. Using several scalable methods we developed for identifying and annotating protein SAE features, we discovered that ESM-2's latent features accurately capture many known biophysical properties - including catalytic active sites, zinc finger domains, mitochondrial targeting sequences, phosphorylated residues, disulfide bonds, and disordered regions. By quantitatively comparing these identified features to known concept annotations, we establish new evaluation metrics for SAEs. We further use Claude-3.5 Sonnet (new) \cite{anthropic_claude_2024} to provide automated annotations of latent features and use PLM feature activation patterns to identify missing and potentially new protein annotations. We show how PLM features can be used to steer model outputs in interpretable ways. To facilitate exploration of these features, we provide an interactive visualization platform \url{interPLM.ai}, along with code for training and analyzing features at \url{github.com/ElanaPearl/interPLM}.

\section{Related Works \& Background}
\subsection{PLM Interpretability}
PLMs are deep learning models (typically transformers) that perform self-supervised training on protein sequences, treating amino acids as tokens in a biological language to learn their underlying relationships and patterns \cite{simon_language_2024}. These models consist of multiple transformer layers, each containing attention mechanisms and neural networks that progressively build up intermediate representations of the protein sequence. Prior work on interpreting PLMs has focused on analyzing these internal representations - both by probing the hidden states at different layers and by examining the patterns of attention between amino acids. Studies have demonstrated that attention maps can reveal protein contacts \cite{rao_transformer_2020} and binding pockets, while the hidden state representations from different layers can be probed to predict structural properties like secondary structure states \cite{vig_bertology_2020}. Recent evidence suggests that rather than learning fundamental protein physics, PLMs primarily learn and store coevolutionary patterns  - coupled sequence patterns preserved through evolution \cite{zhang_protein_2024}. This finding aligns with traditional approaches that explicitly modeled coevolutionary statistics \cite{marks_protein_2011}, and with modern deep learning methods that leverage evolutionary relationships in training \cite{abramson_accurate_2024}.

However, even if PLMs are memorizing evolutionarily conserved patterns in motifs, key questions remain about their internal mechanisms: How do they identify conserved motifs from individual sequences? What percentage of learned features actually focus on these conserved motif patterns? How do they leverage these memorized patterns for accurate sequence predictions? What additional computational strategies support these predictions? Answering these questions could both reveal valuable biological insights and guide future model development.
\subsection{Sparse Autoencoders (SAEs)}
In attempts to reverse-engineer neural networks, researchers often analyze individual neurons - the basic computational units that each output a single activation value in response to input. However, work in mechanistic interpretability has shown that these neurons don't map cleanly to individual concepts, but instead exhibit superposition - where multiple unrelated concepts are encoded by the same neurons \cite{arora_linear_2018,olah_zoom_2020}. Sparse Autoencoders (SAEs) are a dictionary learning approach that addresses this by transforming each neuron's activation into a larger but sparse hidden layer \cite{yun_transformer_2023,cunningham_sparse_2023,bricken_towards_2023}.

At their core, SAEs learn a "dictionary" of sparsely activated features that can reconstruct the original neuron activations. Each feature i is characterized by two components: its dictionary vector ($\mathbf{d}_i$) in the original embedding space (stored as rows in the decoder matrix) and its activation value ($f_i$) that determines its contribution. The reconstruction of an input activation vector $\mathbf{x}$ can be expressed as:
$$ \mathbf{x} \approx \mathbf{b} + \sum_{i=1}^{d_{\text{dict}}} f_i(\mathbf{x})\mathbf{d}_i $$
where $\mathbf{b}$ is a bias term and $d_{\text{dict}}$ is the size of the learned dictionary. This decomposition allows us to represent complex neuron activations as combinations of more interpretable features (see Appendix Figure \ref{fig:sae-diagram} for more details).

SAE analysis has advanced our understanding of how language and vision models process information \cite{cunningham_sparse_2023,gorton_missing_2024}. Neural network behavior can be understood through computational circuits - interconnected neurons that collectively perform specific functions. While traditional circuit analysis uncovers these functional components (like edge detectors \cite{olah_zoom_2020} or word-copying mechanisms \cite{olsson_-context_2022}), using SAE features instead of raw neurons has improved the identification of circuits responsible for complex behaviors \cite{marks_sparse_2024}.

Researchers can characterize these features through multiple approaches: visual analysis \cite{mcdougall_sae_2024}, manual inspection \cite{bricken_towards_2023}, and large language model assistance \cite{bills_language_2023}. Feature functionality can be verified through intervention studies, where adjusting feature activation values steers language model outputs toward specific behaviors, demonstrating their causal role \cite{templeton_scaling_2024}.

One of the field's main challenges lies in evaluation. While technical metrics like how well autoencoders reconstruct their inputs are straightforward, assessing feature interpretability remains subjective. Recent work has made progress by using text-based games, where feature activations can be mapped to labeled game states in chess and Othello \cite{karvonen_measuring_2024}. Protein studies offer similar potential through their structural and functional annotations, though interpreting biological features requires domain expertise that makes evaluation more challenging than in language or vision domains.

Our work with InterPLM demonstrates that while PLM neurons exhibit polysemantic behavior, SAE analysis reveals more interpretable features. We present a comprehensive framework of quantitative metrics and methods for visualizing, analyzing, describing, steering, and learning from these PLM-derived SAE features.

\section{Results}
\begin{figure}[htbp]
    \vspace{-15pt}
    \centering
    \includegraphics[width=\linewidth]{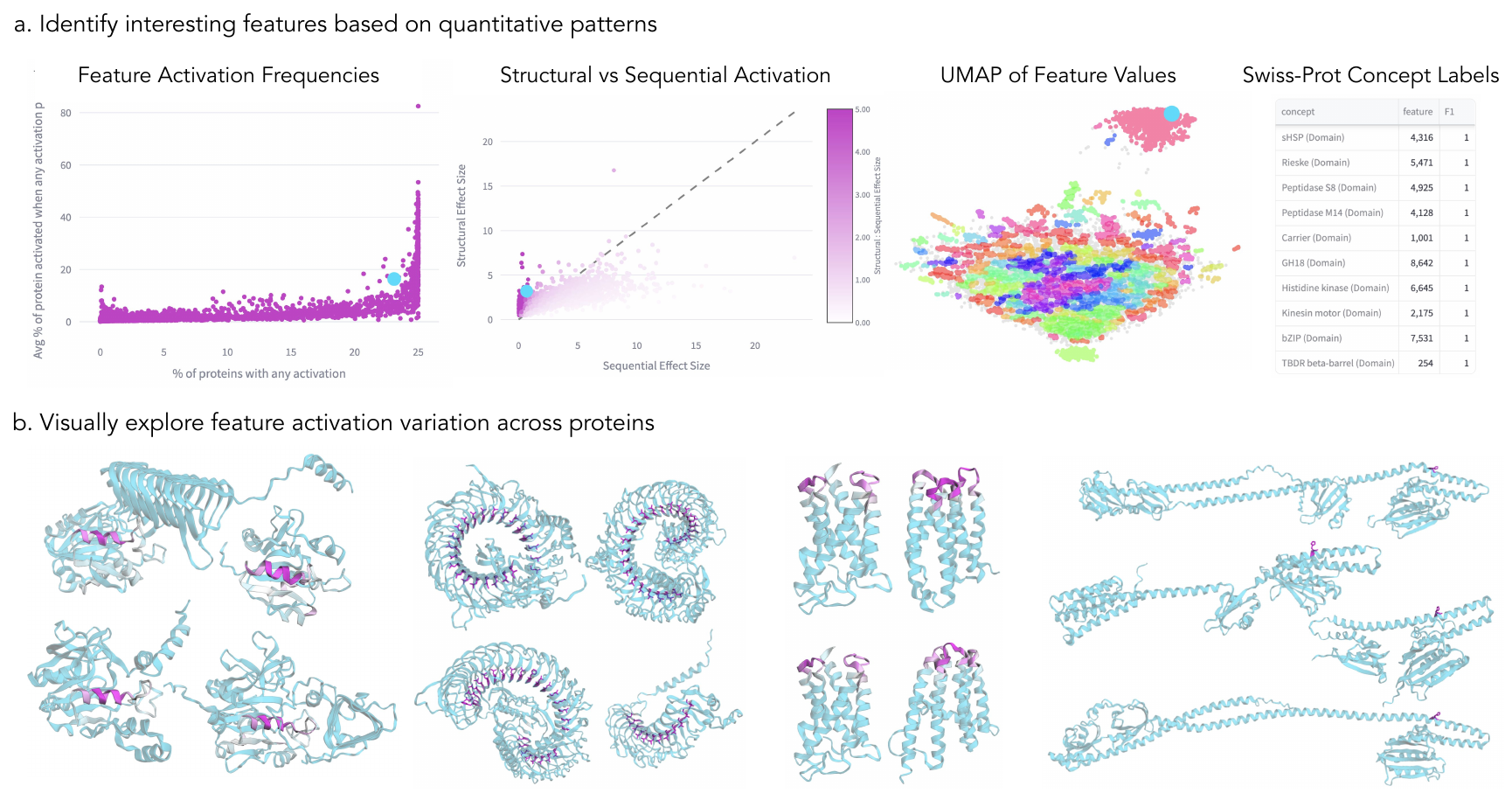}
    \caption{\textbf{SAE feature analysis and visualizations reveal features with diverse and consistent activation patterns.} a) Quantitative comparison of learned features through four complementary approaches: 1) Feature activation frequency distribution showing the relationship between proteome-wide prevalence (x-axis) and protein-specific activation strength (y-axis), revealing both ubiquitous and selective features 2) Structural vs. sequential activation patterns, comparing feature activation strengths in 3D versus sequence proximity to peak activation sites, revealing features that operate through either structural or sequential mechanisms 3) UMAP embedding of feature vectors, illustrating natural clustering of related structural/functional motifs 4) Swiss-Prot concept mapping results, linking learned features to known biological concepts. b) Structural visualization of four representative features (left to right: f/1854, f/10230, f/8144, f/8128) mapped onto example proteins, with each feature highlighted in the figure above it from (a) in blue.}
    \label{fig:dashboard}
\end{figure}
\subsection{Sparse Autoencoders Find Interpretable Concepts in Protein Language Models}
To understand the internal representations of protein language models (PLMs), we trained sparse autoencoders (SAEs) on embeddings from ESM-2-8M. We expanded each of ESM-2's six layers from 320 neurons to 10,420 latent features. The quantity and domain-specificity of these features necessitated developing both qualitative visualization methods and quantitative evaluation approaches to enable interpretation. See \hyperref[methods]{\textbf{Methods}} for training and analysis details.

Analysis of feature activation patterns revealed three distinct modes of amino acid recognition: structural patterns (coordinated activation between spatially or sequentially proximal residues), protein-wide patterns (broad activation across single proteins), and functional patterns (consistent activation within annotated domains, motifs, and binding sites).

When evaluated with our quantitative measures of biological interpretability, we find significantly more conceptual features in SAE features than ESM neurons. We also show that this feature interpretation can be automated with a large language model, and once a feature has an interpretation, this can be used to suggest missing and new protein annotations.

Through clustering features based on their dictionary vectors, we identified groups that detect related biological structures with distinct specializations, from subtle variations in kinase binding site recognition to hierarchical relationships between specific and general beta barrel detectors. We further show that targeted interventions on specific feature values enable controlled manipulation of the model's representations, producing interpretable changes that propagate to influence predictions even for non-intervened amino acids.

For consistency and clarity, our visualizations primarily focus on features from the fourth layer (denoted f/feature-number) unless explicitly stated otherwise.

\subsection{InterPLM: An Interactive Dashboard for Feature Exploration}
Features from SAEs of every ESM-2 layer can be explored through \url{InterPLM.ai}, which provides multiple complementary views of feature behavior: (1) sequential vs structural activation patterns, revealing how features capture both local sequence motifs and long-range structural relationships, (2) protein coverage analysis, distinguishing between features that identify specific local properties versus those capturing broader domain-level patterns, (3) feature similarity through UMAP \cite{mcinnes_umap_2018} visualization, and (4) alignment with known Swiss-Prot \cite{poux_expert_2017} concepts. These feature's characteristic activation modes often persist across diverse proteins, suggesting that many features capture consistent biochemical and structural properties (Figure \ref{fig:dashboard}). By interactively examining the full distribution of features through multiple lenses - their protein coverage, activation patterns, and spatial properties - we uncover a spectrum of behaviors, ranging from precise detectors that activate on single amino acids across many proteins to broader features that respond to complete sequences of specific protein families. As shown in Figure \ref{fig:example_features}, examining the spatial distribution of activation sites uncovers features sensitive to various forms of proximity – whether sequential, structural, or both – providing insights into how the model encodes different aspects of protein organization.
\begin{figure}[htbp]
    \centering
    \includegraphics[width=\linewidth]{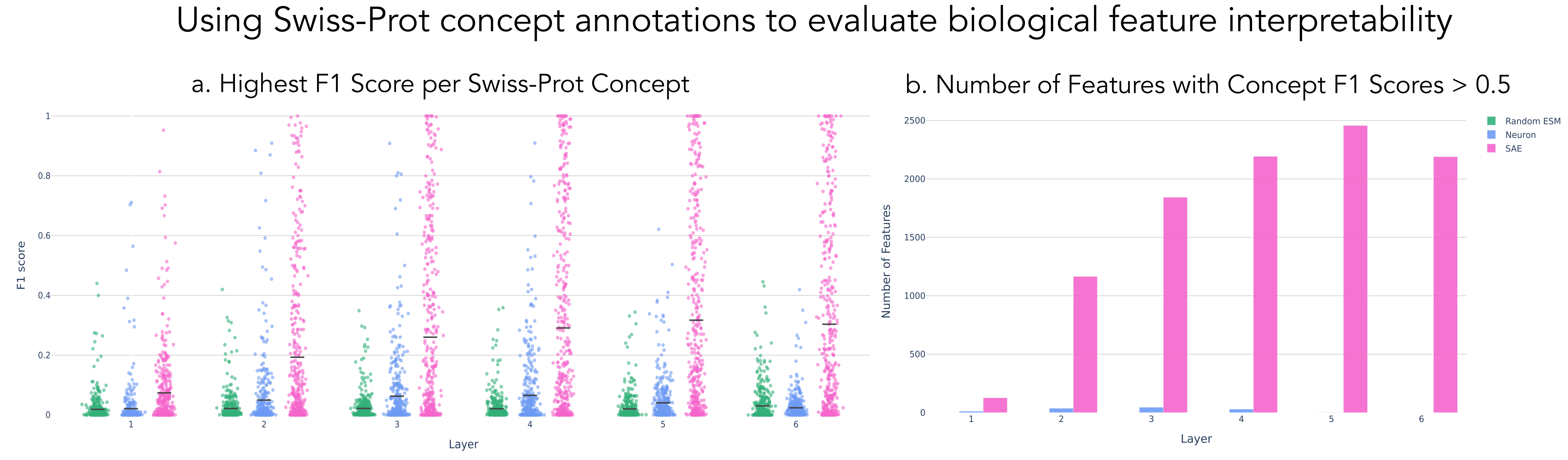}
    \caption{\textbf{SAE features have stronger associations with Swiss-Prot concepts than ESM neurons.}
Comparing the features of an SAE model trained on ESM-2 embeddings (pink), the original neurons of the ESM-2 embeddings (blue), and the features of an SAE trained on embeddings from an ESM-2 model with shuffled weights (green). Models are compared based on the F1 scores between features and Swiss-Prot concepts. 
(a) For each concept, select one feature from the validation set (based on highest F1 with that concept), and visualize the F1 for that feature and concept in a held-out set.
(b) For each layer, count the number of features that have an F1 score with any concept > 0.5 in both the validation and held-out sets.}
    \label{fig:concepts}
\end{figure}
\subsection{Sparse Autoencoder Features Capture More Biological Annotations than Neurons}

When comparing binarized feature activation values to biological concept annotations extracted from Swiss-Prot (dataset and methodological details in \hyperref[methods]{Methods}), we observed that our learned features are much more specific than many concept annotations. The 433 Swiss-Prot concepts we evaluated span structural patterns, biophysical properties, binding sites, and sequential motifs (concept categories enumerated in Appendix \ref{sec:swissprot-metadata}). While some concepts are specific to individual amino acids, others can cover broad domains that include many different physical regions, resulting in standard classification metrics overly penalizing features that activate on individual subsets of larger domains. For example, f/7653 activates on two conserved positions in tyrosine recombinase domains. While this feature has perfect precision on the 45 tyrosine recombinases in our test set, it has a recall of 0.011 because tyrosine recombinase domains are hundreds of amino acids long so it 'misses' most example. To address this, we calculate precision of feature prediction per amino acid, but calculate recall \textit{per domain}. In the case of the tyrosine recombinase feature, because it correctly identifies two amino acids in every domain tested, it now has a domain-adjusted recall of 1.0. Combining these to calculate the F1 scores between feature activations and concepts, we can quantitatively compare interpretability of SAE features versus the original neurons, even though features and concepts are defined at different levels of granularity.

\begin{figure}[htbp]
    \centering
    \includegraphics[width=\linewidth]{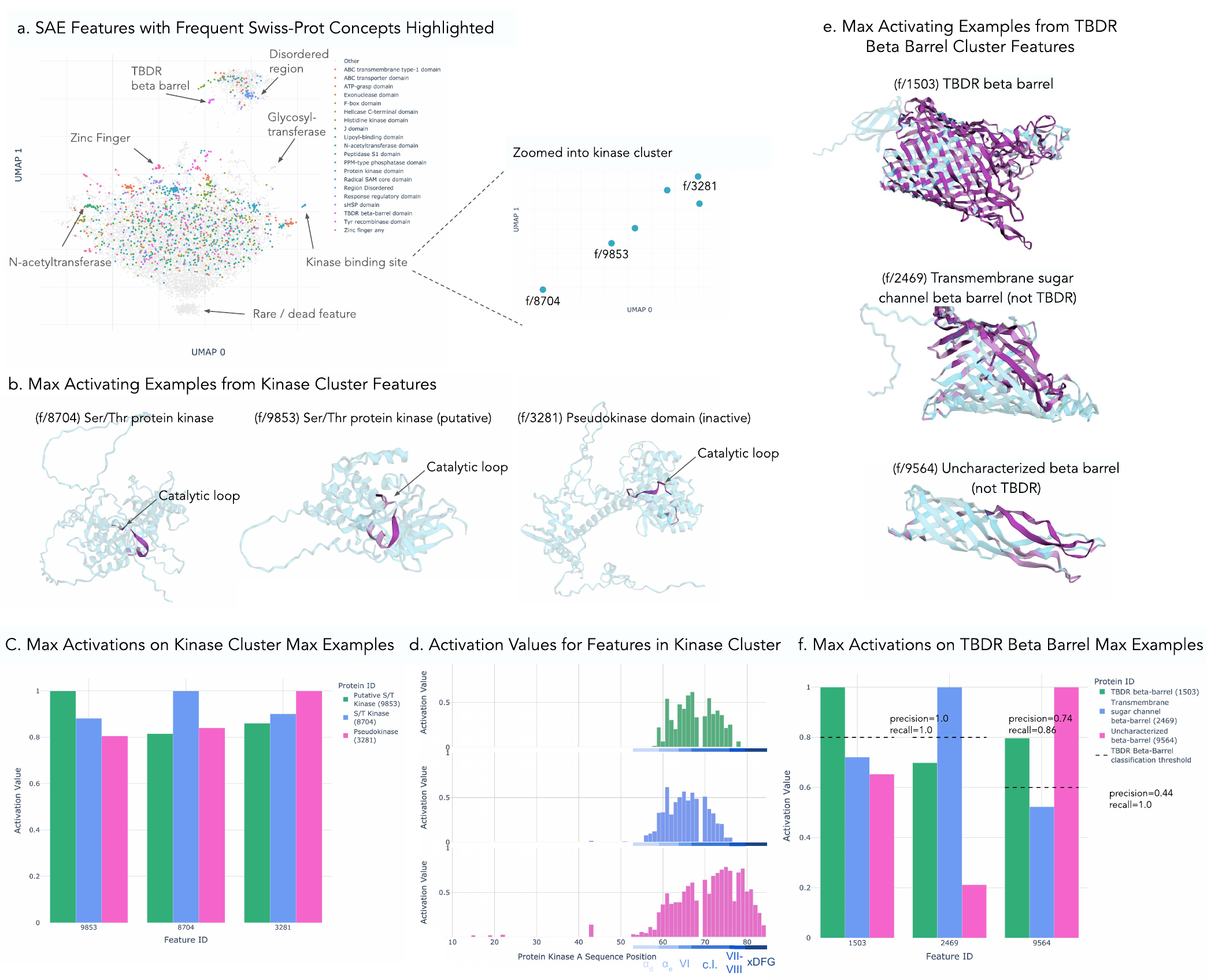}
    \caption{\textbf{Clustering reveals groups of features with similar functional and structural roles but subtle differences in activation patterns.}
    (a) UMAP of SAE features clustered based on their dictionary values. Features associated with one of the top 20 most commonly labeled Swiss-Prot concepts are highlighted in color and severely random clusters (Swiss-Prot labeled or not) are manually identified. In particular, one cluster of kinase features is expanded. (b) Structures of the maximum activating examples from 3 features selected from cluster of kinase-binding-site features. Higher activation values indicated with darker pink and location of catalytic loop specified. (c) Comparing the maximum activation value of each feature on the maximally activating proteins selected by the other kinases. (d) Comparing the physical locations within the kinase binding site these features activate, with specific annotations for each sub-region within kinase binding site that has feature activation. All features visualized on protein kinase A (Uniprot ID: P17612) (e) Structures of maximum activating examples from 3 features selected from cluster of TBDR beta barrels (f) Activation patterns of 3 proteins in a cluster unlabeled by Swiss-Prot concepts but identified as glycosyltransferase cluster. All features visualized on glycosyltransferase amsK (Uniprot: Q46638), the maximally activating protein for f/9047.
}
    \label{fig:clusters}
\end{figure}

While individual neurons showed at most 46 clear concept associations per layer, SAE features revealed up to 2,548 features per layer with strong concept alignment (Figure \ref{fig:concepts}). This difference is explained by both the SAE features' ability to capture more specific qualities within each concept category and their detection of a broader range of concepts overall, expanding from 15 distinct concepts with neurons to 143 with SAE features. Full list of associated Swiss-Prot concepts from features and neurons in Appendix (\ref{tab:protein_concepts_a},\ref{tab:protein_concepts_b}). This dramatic difference persisted across model layers and disappeared in control experiments with randomized model weights. Notably, although the randomized models extract zero features corresponding to biological Swiss-Prot concepts (Figure \ref{fig:concepts}), they have hundreds of features strongly associated with individual amino acid types (Appendix Figure \ref{fig:amino_acid}).

\subsection{Features Form Clusters Based on Shared Functional and Structural Roles}
Clustering features by their dictionary vectors exposes groups with coherent biological functions and distinct specializations. Within a kinase-associated feature cluster, three features exhibited activation on binding site regions with subtle variations in their spatial preferences (Figure \ref{fig:clusters}). One feature activated specifically on the conserved alpha helix and beta sheet ($\alpha_e$, VI) preceding the catalytic loop, while others concentrated on distinct regions near the catalytic loop, as evidenced by their maximally activated examples in both structural and sequence representations (\ref{fig:clusters}b,c). Though their peak activation positions varied, all features maintained high activation levels (> 0.8) on their respective maximum examples across the cluster, suggesting they identify similar kinase subtypes.

Analysis of a beta barrel-associated cluster revealed a distinct generalization pattern. While all three features in the cluster were labeled as TonB-dependent receptor (TBDR) beta barrels, only one feature exhibited specificity for TBDR beta barrels, whereas the other two identified beta barrel structures more broadly, including TBDRs, as demonstrated by their maximally activating examples (\ref{fig:clusters}e). All three features exhibit F1 associations with TBDR beta barrels, yet they differ markedly in their specificity. f/1503 shows exceptional specificity (F1=.998), functioning as a true TBDR detector. The other features, though capable of identifying TBDR structures (high recall), also recognize various other beta barrel proteins (lower precision), resulting in lower but still significant F1 scores (.793, .611). This clustering demonstrates how the model's embedding space captures the natural hierarchy of beta barrel structures, where specialized TBDR detectors and general beta barrel features maintain similar representations despite operating at different levels of structural specificity.
\begin{figure}[htbp]
    \vspace{-5pt}
    \centering
    \includegraphics[width=0.8\linewidth]{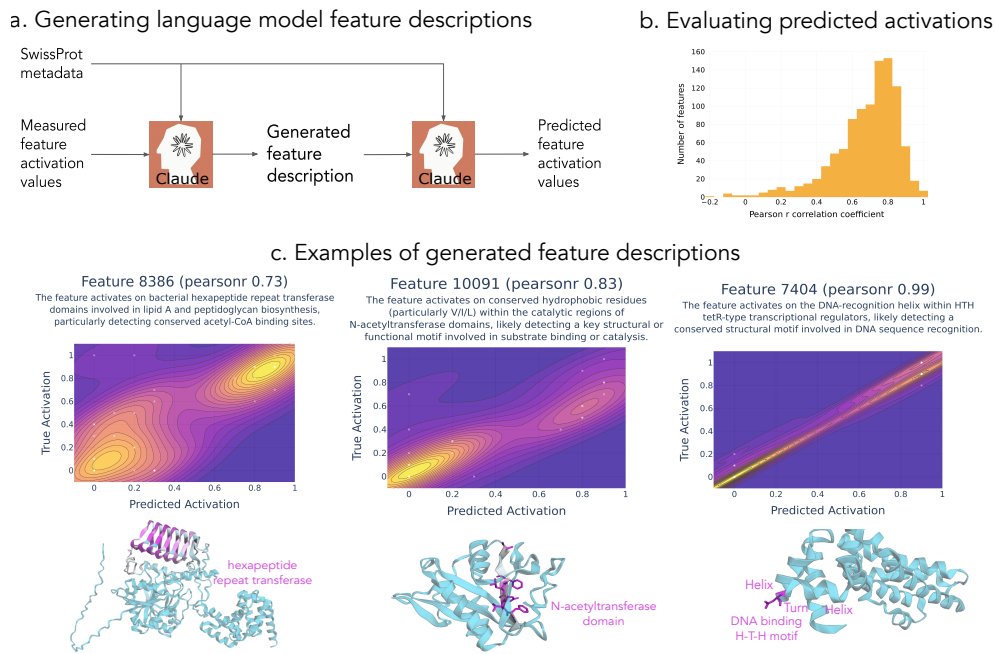}
    \caption{\textbf{Language models can generate automatic feature descriptions for SAE features.}
(a) Workflow for generating and validating descriptions with Claude-3.5 Sonnet (new). (b) Comparing measured maximum activation values in proteins to predicted maximum activation values via Pearson r correlation across 1200 features. (c) Examples of generated feature descriptions and maximally activated proteins of each feature. Predicted activations quality visualized via kernel density estimation. The text is Claude's description summary of each feature. Elements of description present in max examples annotated next to structures.
}
    \label{fig:llm}
\end{figure}
\subsection{Large Language Models Can Generate Meaningful Feature Descriptions}
To extend beyond Swiss-Prot concepts, which label less than 20\% of features across layers (Appendix Figure \ref{fig:pct-covered}), we developed an automated pipeline using Claude to generate feature descriptions. By providing Claude-3.5 Sonnet (new) with the Swiss-Prot concept information including text information not applicable for classification, along with examples of 40 proteins with varying levels of maximum feature activation, we generate descriptions of what protein and amino acid characteristics activate the feature at different levels. As validation, the model generated descriptions and Swiss-Prot metadata are used to predict feature activation levels on separate proteins and showed high correlation with actual feature activation (median Pearson r correlation = 0.72) across diverse proteins. As shown in Figure \ref{fig:llm}, the descriptions accurately match specific highlighted protein features, with density plots revealing distinct clusters of correctly predicted high and low feature activations. The two examples with higher correlations identify specific conserved motifs directly tied to consistent protein structure and function, while the third example (f/8386) detects a structural motif (hexapeptide beta helix) that does not have Swiss-Prot annotations and appears across multiple functionally diverse proteins, potentially explaining the increased difficultly and lower performance at protein-level activation prediction. However, we note that across the distribution of tested features, there is only a weak dependence between a feature's Swiss-Prot concept prediction performance and its LLM description accuracy (Pearson r = 0.11), suggesting that LLM-generated descriptions can effectively characterize many protein features regardless of whether they are easily categorized by annotations (Appendix Figure \ref{fig:f1-v-pearsonr}).

\begin{figure}[htbp]
    \vspace{-2pt}
    \centering
    \includegraphics[width=0.8\linewidth]{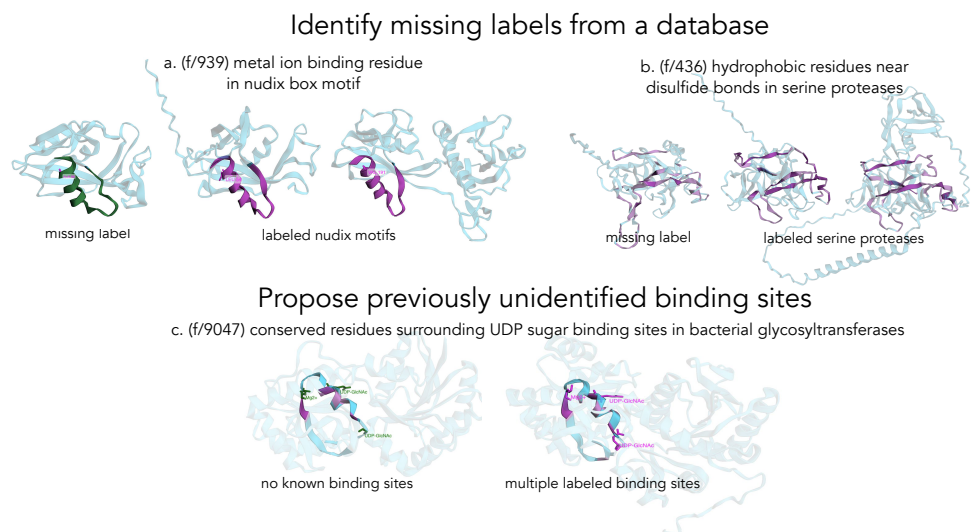}
    \caption{\textbf{Feature activation patterns can be used to identify missing and new protein annotations.}
(a) f/939 identifies missing motif annotation for Nudix box. It activates on a single amino acid in conserved position which is labeled in each structure. Right two proteins are examples with high activations and pink coloring on the region with Nudix label. Left protein (B2GFH1), which does not have a Nudix motif annotation in Swiss-Prot, has implied feature activation highlighted in green. The presence of a Nudix motif somewhere in this protein is confirmed by InterPro. (b) f/436 identifies missing domain annotation for peptidase S1. It activates on a span of ~80 amino acids with shared structural pattern. Right two proteins have high activation on confirmed peptidase S1 domains, with higher activation highlighted in pink. Left protein, which does not have peptidase S1 domain annotation in Swiss-Prot, is likely a missing annotation. The presence of an S1 domain somewhere in this protein is confirmed by InterPro. (c) f/9046 suggests missing binding site annotations for UDP-N-acetyl-$\alpha$-D-glucosamine (UDP-GlcNAc) and Mg2+ within bacterial glycosyltransferases. In both structures, higher activation indicated with darker pink. Right protein has Swiss-Prot annotations for binding sites labeled in pink. Left protein, which has no known binding site annotations, but does have glycosyltransferase activity, has implied binding site annotations labeled in green
}
\vspace{-5pt}
    \label{fig:annotations}
\end{figure}
\subsection{Feature Activations Identify Missing and Novel Protein Annotations}
Analysis of features with strong concept associations revealed cases where "false positive" activations actually indicated missing database annotations. For example, f/939 highlights a single conserved amino acid in a Nudix box motif. We found one example that has high activation of 939 but no Swiss-Prot label, while every other highly activated protein does have the annotation. As seen in Figure \ref{fig:annotations}, the region surrounding this activated position greatly resembles the labeled Nudix motifs. This feature is likely identifying a missing label, and if we examine the InterPro \cite{paysan-lafosse_interpro_2023} entry for this gene, we find reference to a Nudix motif, confirming this assumption. Similarly, f/436 can be used to identify Peptidase S1 domains with missing labels in Swiss-Prot despite positive identification in other external databases and Feature 9046 can be used to suggest the location of UDP and Mg2+ binding sites. The LLM-generated description of f/9046 emphasizes that it identifies UDP-dependent glycosyltransferases, however the majority of activated proteins only have text-based evidence of this, and no labeled annotations for binding sites. Using the binding site annotations from one protein, which each occur on amino acids adjacent to the activated positions, we can suggest potential annotations for the currently unlabeled proteins, although these would require further efforts to validate.

\subsection{Protein Sequence Generation Can be Steered by Activating Interpretable Features}
\begin{figure}[htbp]
    \vspace{-5pt}
    \centering
    \includegraphics[width=\linewidth]{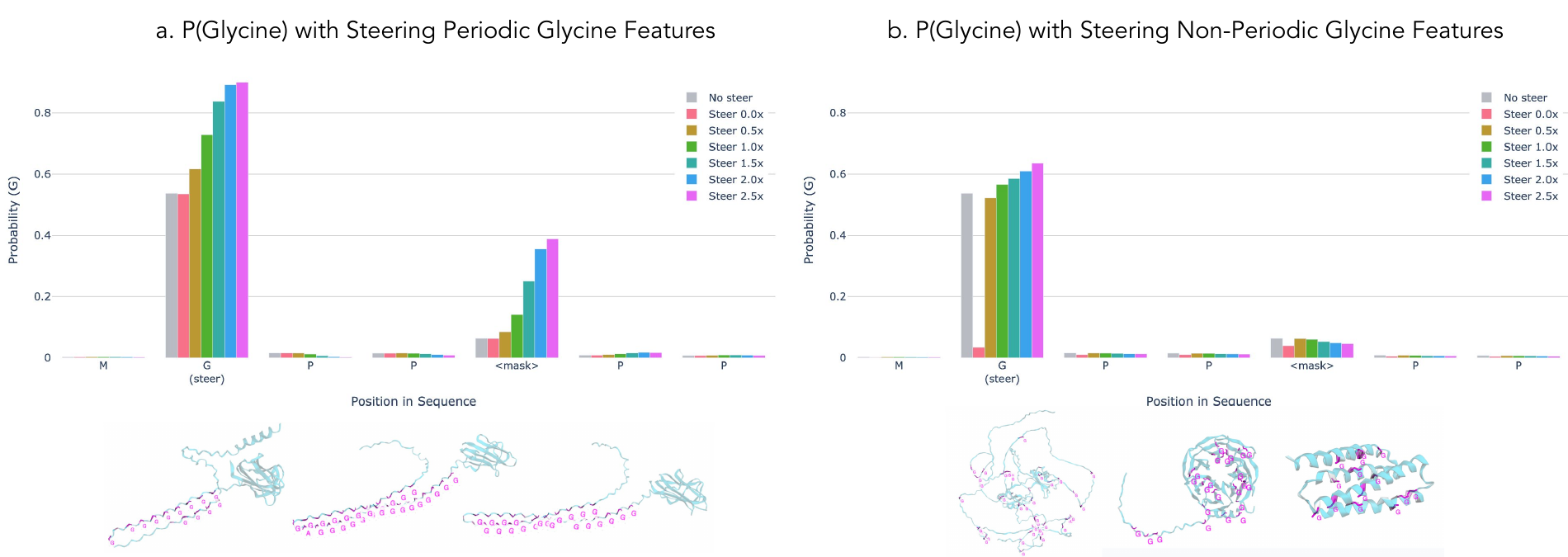}
    \caption{\textbf{Steering feature activation on single amino acid additionally influences protein generation for nearby amino acids in interpretable way for group of periodic glycine features.} Steering activation of features that activate on glycines in periodic repeats of GXX in collagen-like regions by only steering one G and measuring impact on nearby positions.
        (a) Steering periodic glycine features. The result of steering all 3 periodic features on the predicted probability values for glycine (G) at each position in the sequence MGPP<mask>PP. The feature steering was only applied to the unmasked G, not to the masked token or any other positions. Steer amount corresponds to the value the feature was clamped to where 1x is the maximum observed activation value of this feature observed. Maximally activating examples for the 3 Periodic features (f/4616, f/4970, f/10003) shown below (b) Steering non-periodic glycine features. Steering effects and maximally activated examples as in (a) but features selected to have the highest F1 score for glycine (f/6581, f/781, f/5381).
    }
    \label{fig:steer}
\end{figure}
To validate that features capture causally meaningful patterns, we performed targeted interventions in the model's predictions. Unlike in language models where features can be meaningfully steered by clamping their values across entire sequences, protein features rarely maintain consistent activation across domains. This makes it crucial to test whether localized feature manipulation can still influence model behavior at a distance. While many features capture complex biological concepts, quantitatively demonstrating specific interventions' effects is challenging for patterns lacking clear sequence-based validation. We therefore focused on a simple, measurable example: showing how activating specific features can steer glycine predictions in periodic patterns.

We tested three features that activate on periodic glycine repeats (GXXGXX) within collagen-like domains. Given a sequence with a glycine followed by a mask three positions later, increasing these features' activation values on the first glycine position increased the probability of glycine at both that position and the masked position - consistent with the expected periodic pattern (Figure \ref{fig:steer}a). Importantly, these features were originally identified by their characteristic activation pattern occurring every third amino acid in natural sequences. The fact that activating them on just a single glycine position - rather than their standard periodic distribution - still produced interpretable effects demonstrates the robustness of their learned patterns. As expected, highly glycine-specific features (F1 scores: .995, .990, .86) only influenced the directly modified position. In contrast, the periodic glycine features demonstrated a more sophisticated capability: they successfully steered predictions for unmodified positions, even propagating this effect for multiple subsequent periodic repeats with diminishing intensity (Figure \ref{fig:suppl-steer-2}),  revealing their capture of higher-order sequence patterns.

These results demonstrate that features activating on interpretable biological patterns can be capable of causally influence model behavior in predictable ways, even at positions beyond direct manipulation. However, further research is needed to understand the scope and limitations of feature-based steering across different sequence contexts and more complex biological patterns.

\section{Discussion}
Our work demonstrates that sparse autoencoders can extract thousands of interpretable features from protein language model representations. Through quantitative evaluation against Swiss-Prot annotations, we showed that these features capture meaningful biological patterns that exist in superposition within the model's neurons. Our analysis framework combines automated interpretation through large language models with biological database annotations, enabling scalable feature characterization. Beyond identifying features that activate on annotated patterns, we showed that feature activation patterns can identify missing database annotations and enable targeted control of model predictions through interpretable steering.

\subsection{Implications for PLM Interpretability}
The dramatic difference between neuron and SAE feature interpretability (46 vs 2,548 features per layer interpreted with Swiss-Prot concepts) provides strong evidence for information storage in superposition within PLMs. Interestingly, we found that SAEs trained on randomized PLMs still extract features specific to individual amino acids but fail to capture any complex biological concepts we tested for, suggesting that meaningful feature extraction requires learned model weights. This phenomena of identifying amino acids from randomized embeddings aligns with recent observations on randomized language models \cite{templeton_scaling_2024} showing that SAEs capture both properties of the model and the underlying data distribution.

By identifying feature-concept associations through domain-level F1 scores, our biological concept evaluation framework provides a new quantitative approach to assessing interpretability methods, that handles concepts which are annotated more coarsely than the learned features. Furthermore, our LLM-generated feature descriptions achieve strong predictive power at identifying highly activated proteins, even when there are no Swiss-Prot annotations, offering a complementary approach to Swiss-Prot annotations for understanding protein features. While these metrics are only approximations of the true biological interpretability of a model, these metrics clearly distinguish between different approaches and could enable systematic comparison of interpretability techniques. The discovery that "false positive" predictions often indicate missing annotations demonstrates how interpretability tools can provide immediate practical value.

Analysis of feature activation patterns and dictionary vectors reveals both capabilities and open questions about model representations. While we identify many known conserved motifs and 3D patterns, the proportion of features dedicated to motif memorization versus other computational strategies remains unclear. Our framework provides a potential method for quantifying this balance, extending recent work showing PLMs primarily learn coevolutionary statistics \cite{zhang_protein_2024}. Additionally, our steering experiments demonstrate that features can influence both local and neighboring amino acid predictions, though the precise mechanisms by which features contribute to masked token prediction require further investigation.

\subsection{Applications of Interpretable PLM Features}
Our SAE framework offers valuable applications across model development, biological discovery, and protein engineering. For model development, these features enable systematic comparison of learning patterns across different PLMs, revealing how architectural choices influence biological concept capture. Feature tracking during training and fine-tuning could provide insights into knowledge acquisition, while analysis of feature activation during failure modes could highlight systematic biases and guide improvements.

Beyond confirming known patterns, these interpretable features may reveal novel biological insights. Features that don't align with current annotations but show consistent activation patterns across protein families could point to unrecognized biological motifs or relationships that have escaped traditional analysis methods. Additionally, we have demonstrated that feature-based steering can be used to influence sequence generation in targeted ways. While periodic glycine steering may not revolutionize protein engineering itself, it demonstrates a promising new direction for controlling sequence generation during protein design tasks.

\subsection{Limitations and Future Directions}
Scaling this approach to structure-prediction models like ESMFold or AlphaFold represents a crucial next step. Understanding how learned features evolve with increased model size and structural capabilities could enable fine-grained control over generated protein conformations through targeted steering. While we demonstrated successful steering for periodic glycine features in fairly simple contexts, extending steering to more complex biological concepts remains challenging, as we need both robust methods and clear success metrics. While structure prediction models could simplify quantitative evaluation of steering outcomes compared to purely sequence-based models, validating that we've achieved intended biological changes without disrupting other properties is substantially more complex than evaluating semantic shifts in language models.

Feature interpretability could be improved through both recent advances in SAE training methods and additional evaluation metrics beyond Swiss-Prot annotations. Additionally, our analysis currently focuses only on representations of unmasked amino acids, leaving open questions about the information encoded in masked token embeddings and the CLS token, which has proven valuable as a protein-level representation.

While SAE features reveal learned patterns, what remains unexplored is mapping how these features combine into interpretable circuits for specific capabilities – like 3D contact prediction or binding site and allosteric site identification. Such analysis could illuminate how models extract structural and functional information from individual sequences. Finally, systematic analysis of currently uncharacterized features, guided by their activation patterns and impact on predictions, could reveal novel biological motifs – demonstrating how interpretability tools might contribute directly to biological discovery.

\section{Methods}\label{methods}
\subsection{Sparse Autoencoder Training}
\subsubsection{Dataset Preparation}
We selected 5M random protein sequences from UniRef50, part of the training dataset for ESM-2. For each protein, we extracted hidden representations from ESM-2-8M-UR50D after transformer block layers 1 through 6, excluding <cls> and <eos> tokens. The dataset was sharded into groups of 1,000 proteins each, with tokens shuffled within these groups to ensure random sampling during training.

\subsubsection{Architecture and Training Parameters}
We trained sparse autoencoders following the architecture described in \cite{bricken_towards_2023}, using an expansion factor of 32x, creating feature dictionaries of size 10,240 from ESM2-8M embedding vectors of size 320. For each layer, we trained 20 SAEs for 500,000 steps using a batch size of 2,048. Learning rates were sampled in increments of 10x from 1e-4 to 1e-8, with L1 penalties ranging from 0.07 to 0.2. Both the L1 penalty and learning rate were linearly increased from 0 to their final values during the initial training phase, with the learning rate reaching its maximum within the first 5\% of training steps.

\subsubsection{Feature Normalization}
To standardize feature comparisons, we normalized activation values using a scan across 50,000 proteins from Swiss-Prot. For each feature, we identified the maximum activation value across this dataset and used it to scale the encoder weights and decoder weights reciprocally, ensuring all features were scaled between 0 and 1 while preserving the final reconstruction values.

\subsection{Swiss-Prot Concept Evaluation Pipeline}
\subsubsection{Dataset Construction}
From the reviewed subset of UniprotKB (Swiss-Prot), we randomly sampled 50,000 proteins with lengths under 1,024 amino acids. We converted all binary and categorical protein-level annotations into binary amino acid-level annotations, maintaining domain-level relationships for multi-amino-acid annotations. The dataset was split equally into validation and test sets of 25,000 proteins each. We retained only concepts present in either more than 10 unique domains or more than 1,500 amino acids within the validation set.

\subsubsection{Feature-Concept Association Analysis}
For each normalized feature, we created binary feature-on/feature-off labels using activation thresholds of 0, 0.15, 0.5, 0.6, and 0.8. We evaluated feature-concept associations using modified precision and recall metrics:
\begin{align}
\text{precision} &= \frac{\text{TruePositives}}{\text{TruePositives} + \text{FalsePositives}} \\
\text{recall} &= \frac{\text{DomainsWithTruePositive}}{\text{TotalDomains}} \\
\text{F1} &= 2 \cdot \frac{\text{precision} \cdot \text{recall}}{\text{precision} + \text{recall}}
\end{align}
For each feature-concept pair, we selected the threshold yielding the highest F1 score for final evaluation.
\subsubsection{Model Selection and Evaluation}
We conducted initial evaluations on 20\% of the validation set to compare hyperparameter configurations (only including the 135 concepts that had > 10 domains or 1,500 amino acids in this subset alone). For each concept, we identified the feature with the highest F1 score and used the average of these top scores to select the best model per layer. These six models (one per ESM2-8M layer) were used for subsequent analyses and the InterPLM dashboard.

To calculate the test metrics we (1) identify the feature with highest F1 score per-concept on the full validation set, then for each concept, calculate the F1 score of the selected feature on the test set and report these values then (2) identify all feature-concept pairs with F1 > 0.5 in the validation set, calculate their F1 scores on the test set, and report the number of these that have F1 > 0.5 in the test set.

\subsubsection{Baselines}
To train the randomized baseline models, we shuffled the values within each weight and bias of ESM2-8M, calculated embeddings on the same datasets, and repeated the same training (using 6 hyperparameter choices per layer), concept-based model selection, and metric calculation processes. To compare with neurons, we scaled all neuron values between 0 and 1 (based on the minimum and maximum values found in our Swiss-Prot subset), then input these into an SAE with expansion factor of 1x that has an identity matrix for the encoder and decoder, such that all other analysis can be performed identically.

\subsection{LLM Feature Annotation Pipeline}
\subsubsection{Example Selection}
Analysis was performed on a random selection of 1,200 (~10\%) features. For each feature, representative proteins were selected by scanning 50,000 Swiss-Prot proteins to find those with maximum activation levels in distinct ranges. Activation levels were quantized into bins of 0.1 (0-0.1, 0.1-0.2, ..., 0.9-1.0), with two proteins selected per bin that achieved their peak activation in that range, except for the highest bin (0.9-1.0) which received 10 proteins. Additionally, 10 random proteins with zero activation were included to provide negative examples. For features where fewer than 20 proteins reached peak activation in the highest range (0.9-1.0), additional examples were sampled from the second-highest range (0.8-0.9) to achieve a total of 24 proteins between these two bins, split evenly between training and evaluation sets. Features were excluded if fewer than 20 proteins could be found reaching peak activation across the top three activation ranges combined.

\subsubsection{Description Generation and Validation}
For each feature, we compiled a table containing protein metadata, quantized maximum activation values, indices of activated amino acids, and amino acid identities at these positions. Using this data, we prompted Claude-3.5 Sonnet (new) to generate both a detailed description of the feature and a one-sentence summary that could guide activation level prediction for new proteins.

To validate these descriptions, we provided Claude with an independent set of proteins (matched for size and activation distribution) along with their metadata, but without activation information. Claude's predicted activation levels were compared to measured values using Pearson correlation.

\subsection{Feature Analysis and Visualization}
\subsubsection{UMAP Embedding and Clustering}
We performed dimensionality reduction on the normalized SAE decoder weights using UMAP (parameters: metric="euclidean", neighbors=15, min dist=0.1). Clustering was performed using HDBSCAN \cite{hutchison_density-based_2013} (min cluster size=5, min samples=3) for visualization in the InterPLM interface.

\subsubsection{Sequential and Structural Feature Analysis}
We characterized features' sequential and structural properties using the following procedure:
\begin{enumerate}
    \item Identified high-activation regions (>0.6) in proteins with available AlphaFold structures
    \item For each protein's highest-activation residue, calculated:
    \begin{itemize}
        \item Sequential clustering: Mean activation within ±2 positions in sequence
        \item Structural clustering: Mean activation of residues within 6Å in 3D space
    \end{itemize}
   \item Generated null distributions through averaging 5 random permutations per protein
   \item Generated null distributions through averaging 5 random permutations per protein
   \item Assessed clustering significance using paired t-tests and Cohen's d effect sizes across 100 proteins per feature
\end{enumerate}

The 100 proteins per feature were randomly sampled from our Swiss-Prot dataset. Features with < 25 examples meeting this criteria were excluded from this analysis. For structural feature identification in InterPLM, we considered only proteins with Bonferroni-corrected structural p-values < 0.05, with features colored based on the ratio of structural to sequential effect sizes.

\subsection{Steering Experiments}
Following the approach described in \cite{templeton_scaling_2024} we decomposed ESM embeddings into SAE reconstruction predictions and error terms. For sequence steering, we (1) Extracted embeddings at the specified layer (2) Calculated SAE reconstructions and error terms (3) Modified the reconstruction by clamping specified features to desired values (4) Combined modified reconstructions with error terms (5) Allowed normal model processing to continue (6) Extracted logits and calculated softmax probabilities for comparison across steering conditions. Steering experiments all conducted using NNsight \cite{fiotto-kaufman_nnsight_2024}.

\section*{Acknowledgements}
Special thanks to Tom McGrath, Alex Tamkin, Nicholas Joseph, and Zou lab members for helpful discussions and feedback. E.S. is supported by NSF GRFP (DGE-2146755) and J.Z is supported by funding from the CZ Biohub.

\appendix
\section{Additional background on Protein Language Models}
Protein language models (PLMs) adapt techniques from natural language processing to learn representations of protein sequences that capture both structural and functional properties. These models typically use transformer architectures, treating amino acids as discrete tokens. While natural language modeling typically uses an autoregressive training method, many modern PLMs employ masked language modeling objectives similar to BERT, as protein sequences fundamentally exhibit bidirectional dependencies - the folding and function of any given amino acid depends on both N-terminal and C-terminal context.
These models learn impressively rich protein representations through self-supervised training on large sequence databases without requiring structural annotations. The learned embeddings capture hierarchical information ranging from local physicochemical properties to global architectural features. Notably, these representations have proven crucial for protein structure prediction - they serve as the input embeddings for dedicated folding models like ESMFold, and even AlphaFold, while not explicitly a language model, dedicates most of the computational resources in a given prediction to learning protein representations from multiple sequence alignments in a conceptually similar manner.
Many efforts have demonstrated that the representations learned through masked language modeling alone contain remarkable structural and functional information, enabling state-of-the-art performance on tasks ranging from structure prediction to protein engineering. This success appears to stem from the models' ability to capture the underlying patterns in evolutionary sequence data that reflect physical and biological constraints.

\section{Extended SAE Details and Analysis}
\subsection{Additional background}
\begin{figure}[htbp]
    \centering
    \includegraphics[width=\linewidth]{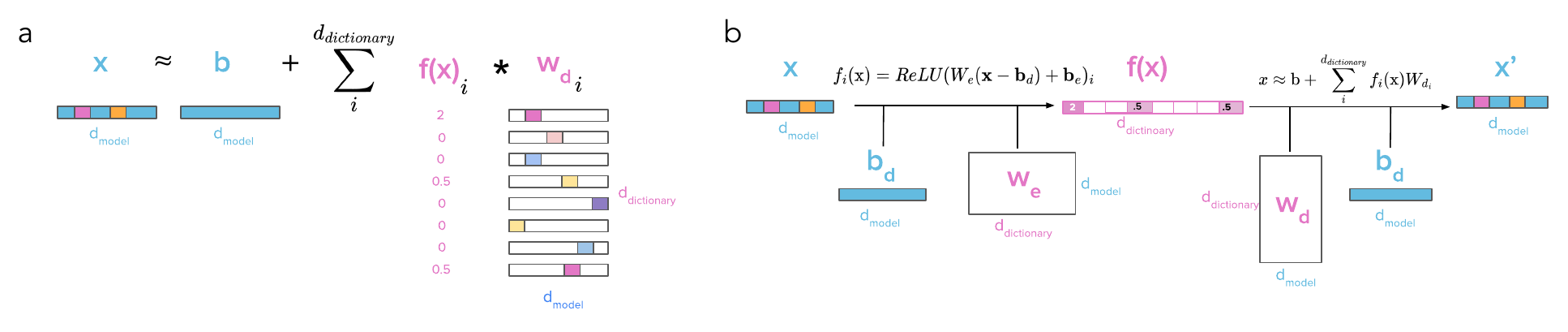}
    \caption{Overview of SAE decomposition and training.\\
    (a) Decomposition of embedding vector into weighted sum of dictionary elements.
(b) Architecture for the SAE}
    \label{fig:sae-diagram}
\end{figure}

SAEs transform the latent vector for a single amino into a new vector with increased size and sparsity. When a specific position in this vector has a non-zero value, that corresponds to the presence of a specific pattern in the amino acid’s neuron embedding. Ideally these model patterns correspond to human interpretable features that we can understand based on patterns by which the feature activates and the impact it has on the model when activated. Specifically, when we perform analysis on feature activation levels, these are using the $f_i(x)i$ values, while dictionary value analysis uses the learned weights in $W_{d_{i}}$ as visualized in \ref{fig:sae-diagram}.

While the SAEs trained in this paper use the architecture above, there are many other alternate SAE methods that vary weight initializations, nonlinearities, loss function, and other training details in aims to simultaneously increase sparsity and increase the reconstruction accuracy \cite{gao_scaling_2024}\cite{rajamanoharan_jumping_2024}\cite{dunefsky_transcoders_2024}.
\subsubsection{SAE metrics}
While the goal of SAE training is to learn features that are maximally interpretable and accurate, these qualities are challenging to explicitly optimize for so during training we optimize for sparsity, hoping that more sparse features are more interpretable, and reconstruction quality.

Specifically, during training we calculate our loss as a weighted sum of an L1 norm calculating the absolute sum of all feature values, and mean squared error calculating how close the reconstructed x' is to the original x. Then, when training SAEs or comparing different implementations, people evaluate L0, the average number of nonzero elements in the latent representation per token and Percent Loss Recovered, the percent of the original cross entropy of the model that is achieved when the model's embeddings are replaced by reconstructions. This last metric measures the amount 'recovered' by comparing the cross entropy of the model using reconstructed embeddings ($CE_{\text{Reconstruction}}$) to the original cross entropy ($CE_{\text{Original}}$) and the cross entropy when the specified embedding layer is instead replaced with all zeros($CE_{\text{Zero}}$) but all later layers remain identical.

Metrics described are calculated per these equations:
\begin{align*}
L_1(f(x)) &= \sum_{i=1}^{d_{\text{dictionary}}} |f_i(x)| \\
MSE(x, x') &= \frac{1}{d_{\text{dictionary}}} \sum_{i=1}^{d_{\text{model}}} (x_i - x'_i)^2 \\
L_0(f(x)) &= \sum_{i=1}^{d_{\text{dictionary}}} \mathbbm{1}(f_i(x) > 0) \\
\text{\% Loss Recovered} &= \Big( 1 - \frac{CE_{\text{Reconstruction}} - CE_{\text{Original}}}{CE_{\text{Zero}} - CE_{\text{Original}}} \Big)* 100
\end{align*}

While we do our final model selection based on Swiss-Prot concept associations as this more closely targets \textit{biological interpretability} than pure sparsity, initial experiments optimized final hyperparameter ranges with these metrics. Final hyperparameters along with $L_0$ and \% Loss Recovered for our 6 SAEs are:

\begin{table}[h]
\centering
\begin{tabular}{|c|c|c|c|}
\hline
Layer & Learning Rate & L1 & \% Loss Recovered \\
\hline
L1 & 1.0e-7 & 0.1 & 99.60921 \\
L2 & 1.0e-7 & 0.08 & 99.31827 \\
L3 & 1.0e-7 & 0.1 & 99.02078 \\
L4 & 1.0e-7 & 0.1 & 98.39785 \\
L5 & 1.0e-7 & 0.1 & 99.32478 \\
L6 & 1.0e-7 & 0.09 & 100 \\
\hline
\end{tabular}
\caption{Layer-wise learning parameters and SAE performance metrics}
\label{table:learning_params}
\end{table}

\section{Swiss-Prot Concept Associations}
\subsection{Swiss-Prot Metadata Categories}
These tables contain all of the metadata used for quantitative feature-concept assoications and LLM feature descriptions and validation, organized into groups with similar themes. The last two columns specify whether each field was used in quantitative concept analysis, LLM descriptions, or both.

\subsection{Swiss-Prot Metadata Categories}
These tables contain all of the metadata used for quantitative feature-concept associations and LLM feature descriptions and validation, organized into groups with similar themes. The last two columns specify whether each field was used in quantitative concept analysis, LLM descriptions, or both.

\subsubsection{Basic Identification Fields}\label{sec:swissprot-metadata}
\begin{longtable}{|>{\raggedright\arraybackslash}p{0.17\textwidth}|>{\raggedright\arraybackslash}p{0.17\textwidth}|>{\raggedright\arraybackslash}p{0.46\textwidth}|>{\centering\arraybackslash}p{0.08\textwidth}|>{\centering\arraybackslash}p{0.08\textwidth}|}
\hline
\textbf{Field Name} & \textbf{Full Name} & \textbf{Description} & \textbf{Quant.} & \textbf{LLM} \\
\hline
\endhead
accession & Accession Number & Unique identifier for the protein entry in UniProt & N & Y \\
\hline
id & UniProt ID & Short mnemonic name for the protein & N & Y \\
\hline 
protein\_name & Protein Name & Full recommended name of the protein & N & Y \\
\hline
gene\_names & Gene Names & Names of the genes encoding the protein & N & Y \\
\hline
sequence & Protein Sequence & Complete amino acid sequence of the protein & N & Y \\
\hline
organism\_name & Organism & Scientific name of the organism the protein is from & N & Y \\
\hline
length & Sequence Length & Total number of amino acids in the protein & N & Y \\
\hline
\end{longtable}

\subsubsection{Structural Features}
\begin{longtable}{|>{\raggedright\arraybackslash}p{0.17\textwidth}|>{\raggedright\arraybackslash}p{0.17\textwidth}|>{\raggedright\arraybackslash}p{0.46\textwidth}|>{\centering\arraybackslash}p{0.08\textwidth}|>{\centering\arraybackslash}p{0.08\textwidth}|}
\hline
\textbf{Field Name} & \textbf{Full Name} & \textbf{Description} & \textbf{Quant.} & \textbf{LLM} \\
\hline
\endhead
ft\_act\_site & Active Sites & Specific amino acids directly involved in the protein's chemical reaction & Y & Y \\
\hline
ft\_binding & Binding Sites & Regions where the protein interacts with other molecules & Y & Y \\
\hline
ft\_disulfid & Disulfide Bonds & Covalent bonds between sulfur atoms that stabilize protein structure & Y & Y \\
\hline
ft\_helix & Helical Regions & Areas where protein forms alpha-helical structures & Y & Y \\
\hline
ft\_turn & Turns & Regions where protein chain changes direction & Y & Y \\
\hline
ft\_strand & Beta Strands & Regions forming sheet-like structural elements & Y & Y \\
\hline
ft\_coiled & Coiled Coil Regions & Areas where multiple helices intertwine & Y & Y \\
\hline
ft\_non\_std & Non-standard Residues & Non-standard amino acids in the protein & N & Y \\
\hline
ft\_transmem & Transmembrane Regions & Regions that span cellular membranes & N & Y \\
\hline
ft\_intramem & Intramembrane Regions & Regions located within membranes & N & Y \\
\hline
\end{longtable}

\subsubsection{Modifications and Chemical Features}
\begin{longtable}{|>{\raggedright\arraybackslash}p{0.17\textwidth}|>{\raggedright\arraybackslash}p{0.17\textwidth}|>{\raggedright\arraybackslash}p{0.46\textwidth}|>{\centering\arraybackslash}p{0.08\textwidth}|>{\centering\arraybackslash}p{0.08\textwidth}|}
\hline
\textbf{Field Name} & \textbf{Full Name} & \textbf{Description} & \textbf{Quant.} & \textbf{LLM} \\
\hline
\endhead
ft\_carbohyd & Carbohydrate Modifications & Locations where sugar groups are attached to the protein & Y & Y \\
\hline
ft\_lipid & Lipid Modifications & Sites where lipid molecules are attached to the protein & Y & Y \\
\hline
ft\_mod\_res & Modified Residues & Amino acids that undergo post-translational modifications & Y & Y \\
\hline
cc\_cofactor & Cofactor Information & Non-protein molecules required for protein function & N & Y \\
\hline
\end{longtable}

\subsubsection{Targeting and Localization}
\begin{longtable}{|>{\raggedright\arraybackslash}p{0.17\textwidth}|>{\raggedright\arraybackslash}p{0.17\textwidth}|>{\raggedright\arraybackslash}p{0.46\textwidth}|>{\centering\arraybackslash}p{0.08\textwidth}|>{\centering\arraybackslash}p{0.08\textwidth}|}
\hline
\textbf{Field Name} & \textbf{Full Name} & \textbf{Description} & \textbf{Quant.} & \textbf{LLM} \\
\hline
\endhead
ft\_signal & Signal Peptide & Sequence that directs protein trafficking in the cell & Y & Y \\
\hline
ft\_transit & Transit Peptide & Sequence guiding proteins to specific cellular compartments & Y & Y \\
\hline
\end{longtable}

\subsubsection{Functional Domains and Regions}
\begin{longtable}{|>{\raggedright\arraybackslash}p{0.17\textwidth}|>{\raggedright\arraybackslash}p{0.17\textwidth}|>{\raggedright\arraybackslash}p{0.46\textwidth}|>{\centering\arraybackslash}p{0.08\textwidth}|>{\centering\arraybackslash}p{0.08\textwidth}|}
\hline
\textbf{Field Name} & \textbf{Full Name} & \textbf{Description} & \textbf{Quant.} & \textbf{LLM} \\
\hline
\endhead
ft\_compbias & Compositionally Biased Regions & Sequences with unusual amino acid distributions & Y & Y \\
\hline
ft\_domain & Protein Domains & Distinct functional or structural protein units & Y & Y \\
\hline
ft\_motif & Short Motifs & Small functionally important amino acid patterns & Y & Y \\
\hline
ft\_region & Regions of Interest & Areas with specific biological significance & Y & Y \\
\hline
ft\_zn\_fing & Zinc Finger Regions & DNA-binding structural motifs containing zinc & Y & Y \\
\hline
ft\_dna\_bind & DNA Binding Regions & Regions that interact with DNA & N & Y \\
\hline
ft\_repeat & Repeated Regions & Repeated sequence motifs within the protein & N & Y \\
\hline
cc\_domain & Domain Commentary & General information about functional protein units & N & Y \\
\hline
\end{longtable}

\subsubsection{Functional Annotations}
\begin{longtable}{|>{\raggedright\arraybackslash}p{0.17\textwidth}|>{\raggedright\arraybackslash}p{0.17\textwidth}|>{\raggedright\arraybackslash}p{0.46\textwidth}|>{\centering\arraybackslash}p{0.08\textwidth}|>{\centering\arraybackslash}p{0.08\textwidth}|}
\hline
\textbf{Field Name} & \textbf{Full Name} & \textbf{Description} & \textbf{Quant.} & \textbf{LLM} \\
\hline
\endhead
cc\_catalytic\_activity & Catalytic Activity & Description of the chemical reaction(s) performed by the protein & N & Y \\
\hline
ec & Enzyme Commission Number & Enzyme Commission number for categorizing enzyme-catalyzed reactions & N & Y \\
\hline
cc\_activity\_regulation & Activity Regulation & Information about how the protein's activity is controlled & N & Y \\
\hline
cc\_function & Function & General description of the protein's biological role & N & Y \\
\hline
protein\_families & Protein Families & Classification of the protein into functional/evolutionary groups & N & Y \\
\hline
go\_f & Gene Ontology Function & Gene Ontology terms describing molecular functions & N & Y \\
\hline
\end{longtable}

\begin{table}[htbp]
\centering
\small
\begin{minipage}[t]{0.32\textwidth}
\textbf{Active Site}
\begin{itemize}
    \item Acyl-ester intermediate
    \item O-(3'-phospho-DNA)-tyrosine intermediate
    \item Tele-phosphohistidine intermediate
\end{itemize}

\textbf{Coiled Coil}
\end{minipage}
\hfill
\begin{minipage}[t]{0.32\textwidth}
\textbf{Compositional Bias}
\begin{itemize}
    \item Acidic residues
    \item Pro residues
\end{itemize}

\textbf{Disulfide Bond}\\

\textbf{Modified Residue}
\begin{itemize}
    \item 4-aspartylphosphate
    \item O-(pantetheine \\  4'-phosphoryl)serine
\end{itemize}
\end{minipage}
\hfill
\begin{minipage}[t]{0.32\textwidth}
\textbf{Signal Peptide}
\begin{itemize}
    \item Tat-type signal
    \item any
\end{itemize}

\textbf{Transit Peptide}
\begin{itemize}
    \item Mitochondrion
    \item any
\end{itemize}

\textbf{Zinc Finger}
\begin{itemize}
    \item CR-type
    \item PHD-type
    \item RING-type
    \item any
\end{itemize}
\end{minipage}
\vspace{1em}

\textbf{Domain [FT]}
\vspace{-5pt}
\begin{multicols}{3}
\begin{itemize}
    \item 2Fe-2S ferredoxin-type
    \item 4Fe-4S ferredoxin-type 1
    \item 4Fe-4S ferredoxin-type 2
    \item AB hydrolase-1
    \item ABC transmembrane type-1*
    \item ABC transporter*
    \item ATP-grasp
    \item C-type lectin
    \item C2
    \item CBS 1
    \item CBS 2
    \item CN hydrolase
    \item CP-type G
    \item Carrier
    \item CheB-type methylesterase
    \item Core-binding (CB)
    \item DPCK
    \item DRBM
    \item Disintegrin
    \item EngA-type G 2
    \item EngB-type G
    \item Era-type G
    \item Exonuclease
    \item F-box*
    \item FAD-binding FR-type
    \item FAD-binding PCMH-type
    \item Fe2OG dioxygenase
    \item Fibronectin type-III 1
    \item Fibronectin type-III 2
    \item G-alpha
    \item GH16
    \item GH18
    \item GMPS ATP-PPase
    \item GST C-terminal
    \item GST N-terminal
    \item Glutamine amidotransferase type-1*
    \item HD
    \item HTH araC/xylS-type*
    \item HTH cro/C1-type
    \item HTH luxR-type
    \item HTH lysR-type
    \item HTH marR-type
    \item HTH tetR-type
    \item Helicase ATP-binding
    \item Helicase C-terminal
    \item Histidine kinase
    \item Ig-like
    \item J*
    \item KH
    \item KH type-2
    \item Kinesin motor
    \item LIM zinc-binding 1
    \item LIM zinc-binding 2
    \item Lipoyl-binding
    \item MPN
    \item MTTase N-terminal
    \item N-acetyltransferase*
    \item NodB homology
    \item Nudix hydrolase
    \item Obg
    \item PDZ
    \item PH
    \item PPIase FKBP-type
    \item PPM-type phosphatase
    \item Peptidase A1
    \item Peptidase M12B
    \item Peptidase M14
    \item Peptidase S1*
    \item Peptidase S8*
    \item Protein kinase*
    \item RNase H type-1
    \item Radical SAM core*
    \item Response regulatory*
    \item Rhodanese
    \item Rieske
    \item S1 motif
    \item S1-like
    \item SH3
    \item SIS
    \item Sigma-54 factor interaction
    \item SpoVT-AbrB 1
    \item SpoVT-AbrB 2
    \item TBDR beta-barrel
    \item TGS
    \item TIR
    \item Thioredoxin
    \item TrmE-type G
    \item Tyr recombinase*
    \item Tyrosine-protein phosphatase
    \item Urease
    \item VWFA
    \item YjeF N-terminal
    \item YrdC-like
    \item bHLH
    \item bZIP
    \item sHSP
    \item tr-type G
\end{itemize}
\end{multicols}
\caption{Swiss-Prot Concepts associated with SAE features in any layer (Part A).
* Indicates concept that is also associated with a neuron in any layer.}
\label{tab:protein_concepts_a}
\end{table}

\begin{table}[htbp]
\centering
\small
\textbf{Motif}
\begin{multicols}{3}
\begin{itemize}
    \item Beta-hairpin
    \item DEAD box
    \item Effector region
    \item Histidine box-2
    \item JAMM motif
    \item NPA 1
    \item Nudix box
    \item PP-loop motif
    \item Q motif
    \item Selectivity filter
\end{itemize}
\end{multicols}

\vspace{1em}

\textbf{Region}
\begin{multicols}{2}
\begin{itemize}
    \item 3-hydroxyacyl-CoA dehydrogenase
    \item A
    \item Adenylyl removase
    \item Adenylyl transferase
    \item Basic motif
    \item Disordered*
    \item Domain II
    \item FAD-dependent cmnm(5)s(2)U34 oxidoreductase
    \item Framework-3
    \item Interaction with substrate tRNA
    \item Large ATPase domain (RuvB-L)
    \item N-acetyltransferase
    \item NBD2
    \item NMP
    \item Pyrophosphorylase
    \item Ribokinase
    \item Small ATPAse domain (RuvB-S)
    \item Uridylyl-removing
    \item Uridylyltransferase
\end{itemize}
\end{multicols}
\caption{Swiss-Prot Concepts associated with SAE features in any layer (Part B).
* Indicates concept that is also associated with a neuron in any layer.}
\label{tab:protein_concepts_b}
\end{table}

\subsection{Additional Concept Coverage Statistics}
\begin{figure}[htbp]
    \centering
    \begin{subfigure}[b]{0.48\linewidth}
        \centering
        \includegraphics[width=\linewidth]{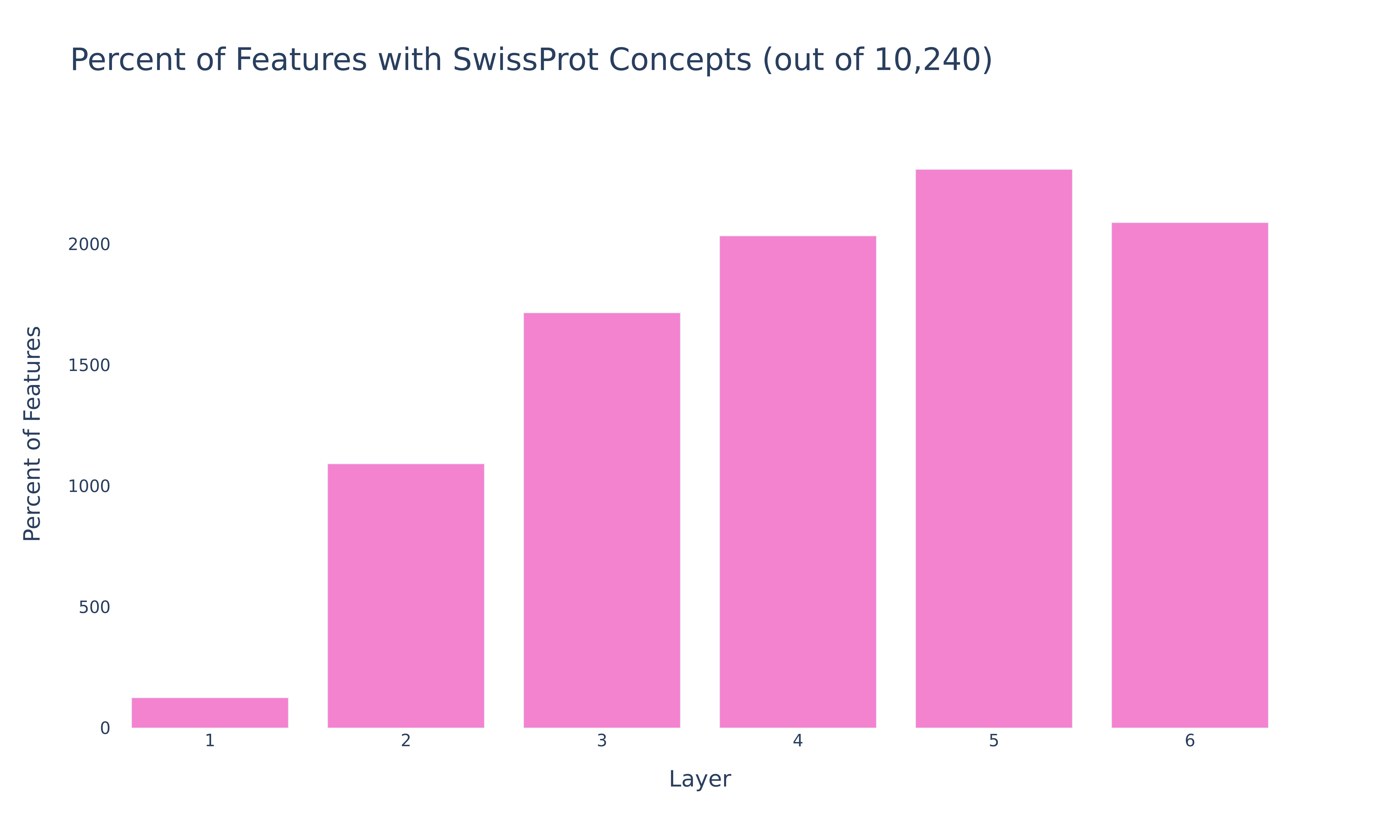}
        \caption{Percent of SAE features per layer that are associated with any Swiss-Prot concept with F1 > 0.5}
        \label{fig:pct-covered}
    \end{subfigure}
    \hfill
    \begin{subfigure}[b]{0.48\linewidth}
        \centering
        \includegraphics[width=\linewidth]{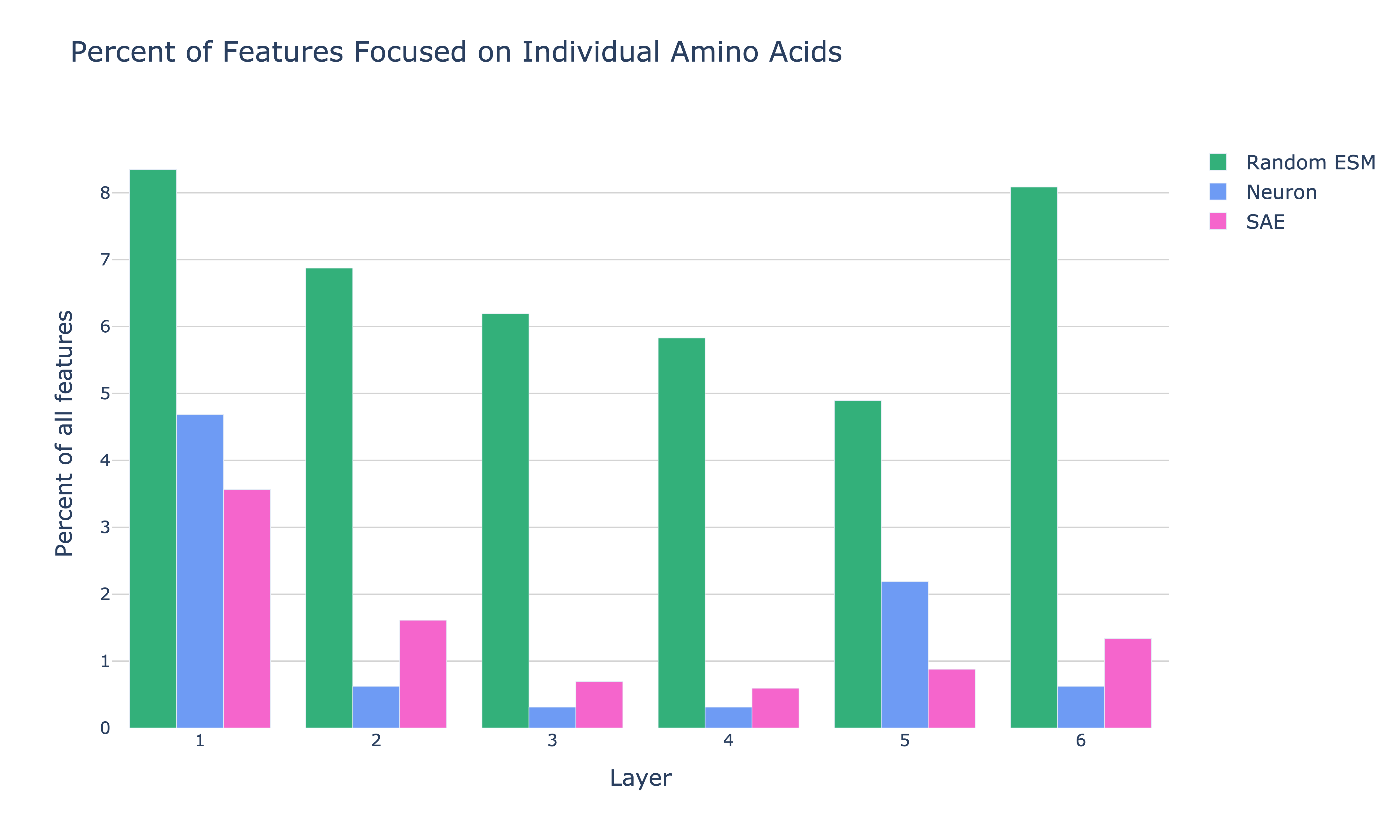}
        \caption{Percent of features (or neurons) in each layer with F1 > 0.5 to an individual amino acid type.}
        \label{fig:pct-covered-b}
    \end{subfigure}
    \caption{Additional feature-concept analysis across layers}
    \label{fig:amino_acid}
\end{figure}

\section{LLM-based Autointerpretability}
\subsection{Prompts}
\begin{tcolorbox}[
    colback=blue!2!white,
    colframe=blue!40!black,
    title={\textbf{Generate description and summary}}
    ]
    Analyze this protein dataset to determine what predicts the 'Maximum activation value' and ‘Amino acids of highest activated indices in protein’ columns. This description should be as concise as possible but sufficient to predict these two columns on held-out data given only the description and the rest of the protein metadata provided. The feature could be specific to a protein family, a structural motif, a sequence motif, a functional role, etc. These WILL be used to predict how much unseen proteins are activated by the feature so only highlight relevant factors for this.\\
    Focus on:\\
    \begin{itemize}
        \item Properties of proteins from the metadata that are associated with high vs medium vs low activation.
        \item Where in the protein sequence activation occurs (in relation to the protein sequence, length, structure, or other properties)
        \item What functional annotations (binding sites, domains, etc.) and amino acids are present at or near the activated positions
        \item  This description that will be used to help predict missing activation values should start with “The activation patterns are characterized by:”
    \end{itemize}
    Then, in 1 sentence, summarize what biological feature or pattern this neural network activation is detecting. This concise summary should start with “The feature activates on”\\
    Protein record: \texttt{\colorbox{blue!10}{Insert table with Swiss-Prot metadata and activation levels}}
\end{tcolorbox}
\begin{tcolorbox}[
    colback=blue!2!white,
    colframe=blue!40!black,
    title={\textbf{Predict activation levels}}
    ]
    Given this protein metadata record, feature description, and empty table with query proteins, fill out the query table indicating the maximum feature activation value within in each protein (0.0-1.0).\\
    
    Base activation value on how well the protein matches the described patterns. There could be 0, 1 or multiple separate instances of activation in a protein and each activation could span 1 or many amino acids.\\
    Output only these values in the provided table starting with "Entry,Maximum activation value". Respond with nothing but this table.\\
    Protein record: \texttt{\colorbox{blue!10}{Insert table with Swiss-Prot metadata}}\\
    Table to fill out with query proteins: \texttt{\colorbox{blue!10}{Insert empty table of IDs to fill out with predictions}} \\
    The activation patterns are characterized by: \texttt{\colorbox{blue!10}{Insert LLM description}}\\
\end{tcolorbox}

\subsection{Example Descriptions and Summaries}

\begin{tcolorbox}[
    colback=blue!2!white,
    colframe=blue!40!black,
    title={\textbf{Full Description} Layer 4 Feature 9047}
    ]
    The activation patterns are characterized by:
    \begin{itemize}
        \item Highest activations (0.9-1.0) occur in glycosyltransferase proteins, particularly glycogen synthases (GlgA) and similar enzymes that transfer sugar molecules
        \item Activated positions consistently occur around amino acid positions 280-450 in these proteins, specifically involving glycine (G), alanine (A), or proline (P) residues
        \item The activated sites frequently overlap with substrate binding regions, particularly nucleotide-sugar binding sites (e.g., ADP-glucose, UDP-glucose, GDP-mannose)
        \item Medium activations (0.3-0.8) are seen in other transferases and synthetases with similar substrate binding patterns
        \item Proteins without sugar/nucleotide binding functions show no activation (0.0)
    \end{itemize}
\end{tcolorbox}

\begin{table}[h]
\centering
\begin{tabular}{|c|c|p{10cm}|}
\hline
Feature & Pearson r & Feature Summary \\
\hline
4360 & 0.75 & The feature activates on interchain disulfide bonds and surrounding hydrophobic residues in serine proteases, particularly those involved in venom and blood coagulation pathways. \\
\hline
9390 & 0.98 & The feature activates on the conserved Nudix box motif of Nudix hydrolase enzymes, particularly detecting the metal ion binding residues that are essential for their nucleotide pyrophosphatase activity. \\
\hline
3147 & 0.70 & The feature activates on conserved leucine and cysteine residues that occur in leucine-rich repeat domains and metal-binding structural motifs, particularly those involved in protein-protein interactions and signaling. \\
\hline
4616 & 0.76 & The feature activates on conserved glycine residues in structured regions, with highest sensitivity to the characteristic glycine-containing repeats of collagens and GTP-binding motifs. \\
\hline
8704 & 0.75 & The feature activates on conserved catalytic motifs in protein kinase active sites, particularly detecting the proton acceptor residues and surrounding amino acids involved in phosphotransfer reactions. \\
\hline
9047 & 0.80 & The feature activates on conserved glycine/alanine/proline residues within the nucleotide-sugar binding domains of glycosyltransferases, particularly at positions known to interact with the sugar-nucleotide donor substrate. \\
\hline
10091 & 0.83 & The feature activates on conserved hydrophobic residues (particularly V/I/L) within the catalytic regions of N-acetyltransferase domains, likely detecting a key structural or functional motif involved in substrate binding or catalysis. \\
\hline
1503 & 0.73 & The feature activates on extracellular substrate binding loops of TonB-dependent outer membrane transporters, particularly those involved in nutrient uptake. \\
\hline
2469 & 0.85 & The feature activates on conserved structural and sequence elements in bacterial outer membrane beta-barrel proteins, particularly around substrate binding and ion coordination sites in porins and TonB-dependent receptors. \\
\hline
\end{tabular}
\caption{Example feature description summaries and their corresponding Pearson correlation coefficients when used along with more verbose descriptions to predict maximum activation levels.}
\label{table:features}
\end{table}

\begin{figure}[htbp]
    \centering
    \includegraphics[width=\linewidth]{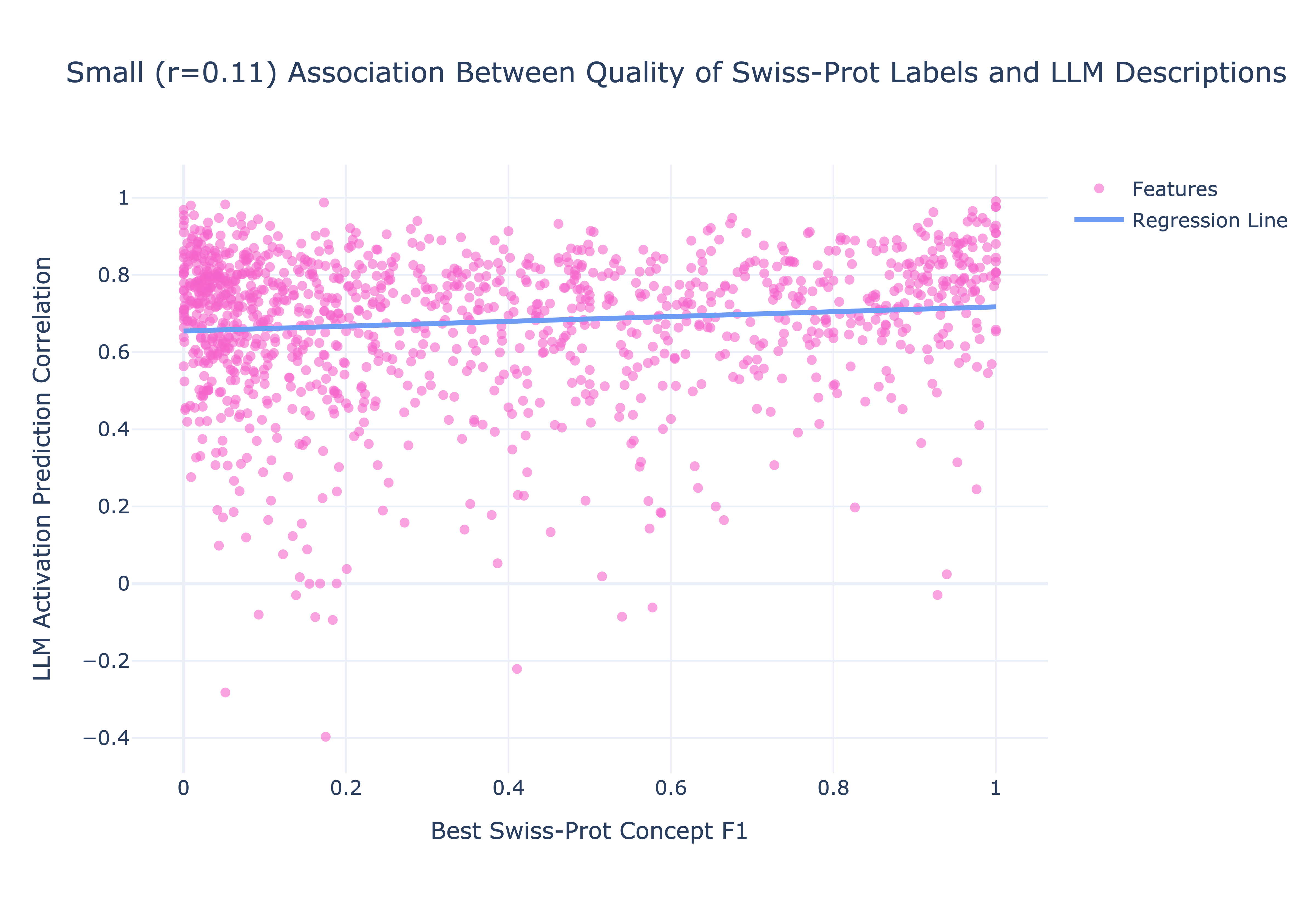}
    \caption{Comparing quality of Swiss-Prot concept labels and accuracy of LLM predicted feature activation patterns reveals small correlation.}
    \label{fig:f1-v-pearsonr}
\end{figure}

\section{Additional cluster example}
\begin{figure}[htbp]
    \centering
    \includegraphics[width=\linewidth]{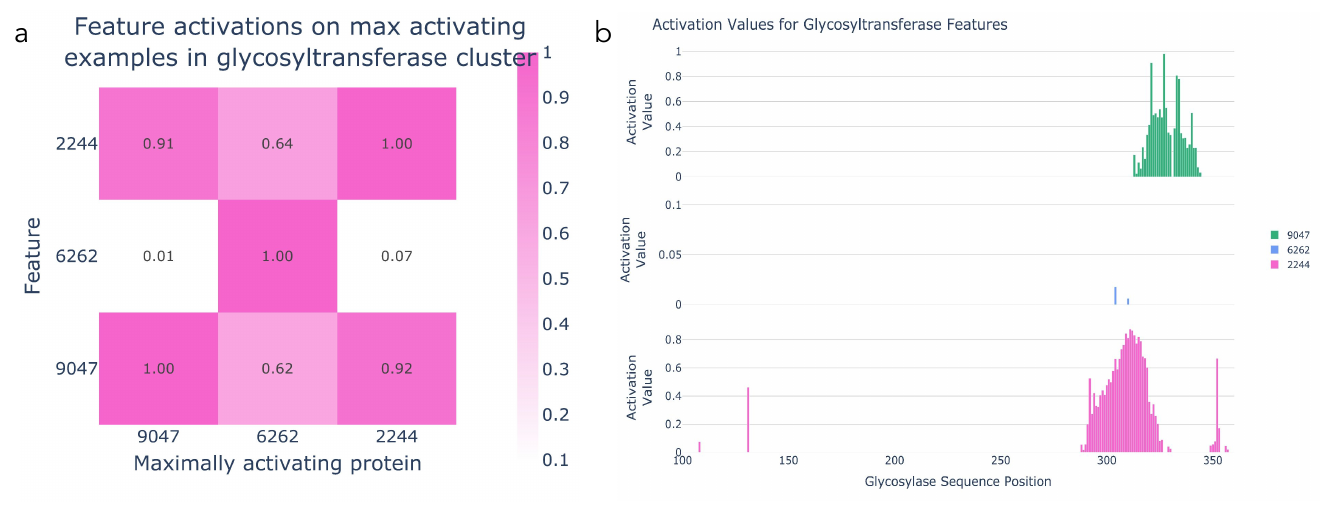}
    \caption{Example of features in glycosylastransferase cluster (2244,6262,9047) with varied maximally activating examples and maximally activating positions (a) Three glycosyltransferase features (y-axis) evaluated on the maximally activating proteins from each of these features (x-axis) (b) All features visualized on glycosyltransferase amsK (Uniprot: Q46638), maximally activating protein for 9047.}
    \label{fig:suppl-glyc-cluster}
\end{figure}

In \ref{fig:suppl-glyc-cluster} we see another example of proteins in common cluster united by a shared conceptual theme but varying in the specific proteins maximally activated and locations. Features 2244, 6262, and 9047 are not identified by a Swiss-Prot concept, but are maximally activated by specific sites within glycosyltransferases (labeled by examining examples at different activation thresholds and Claude-generated descriptions). However, one feature (6262) has minimal activation on the other two maximally activated examples, demonstrating variations in the \textit{type} of glycosylrtansferase identified and also an asymmetry between 6262 appearing more selective in which glycosyltransferases activate it. While the other two features activate on all examples, but particularly on both respective maximally activating examples, when both compared on a separate protein, we see the specific locations within this protein activated are nearby but focused on different regions.

\section{Additional steering experiments}
\begin{figure}[htbp]
    \centering
    \includegraphics[width=\linewidth]{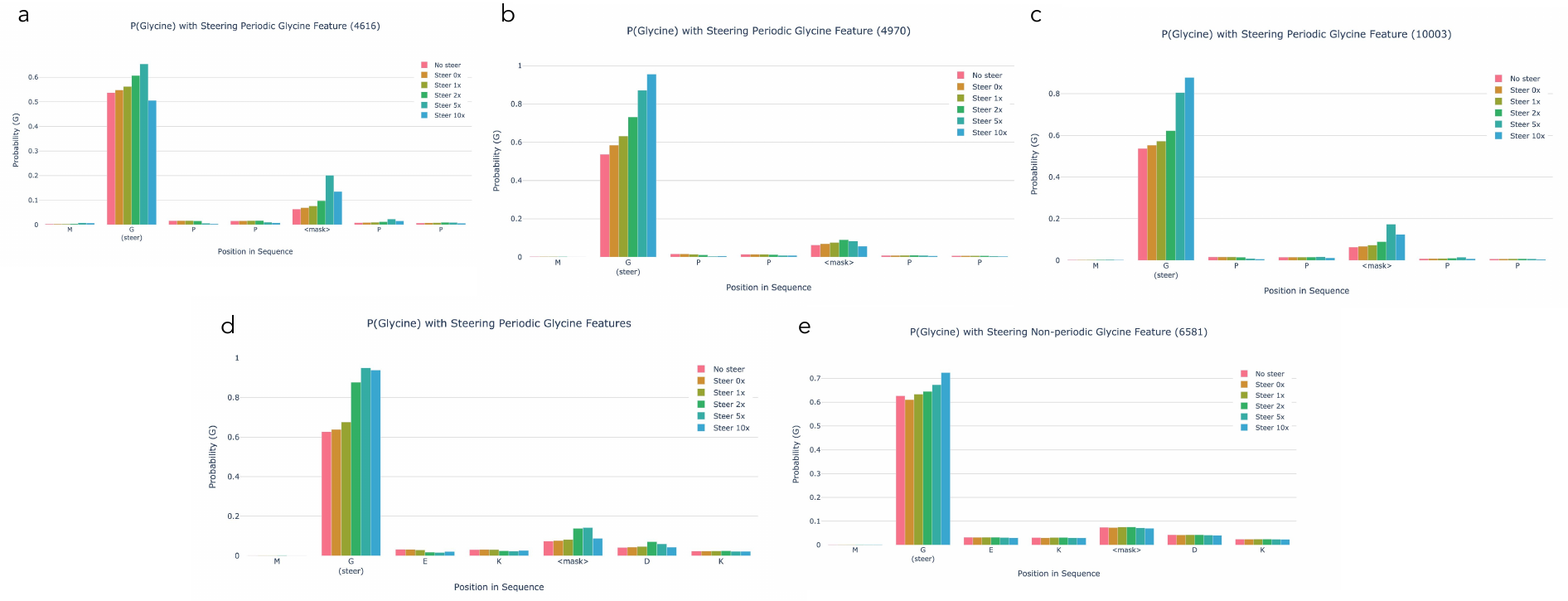}
    \caption{Top row (a-c): Steering MGPP<mask>PP on each of the individual Periodic Glycine Features. Bottom row: Steering an alternate sequence, MGEK<mask>DK. (d) Steering all periodic glycine features (e) Steering non-periodic Glycine feature}
    \label{fig:suppl-steer-1.}
\end{figure}

In Figure \ref{fig:suppl-steer-1.}, we see that each periodic glycine feature can be used to steer the model to predict a periodic glycine pattern. We also evaluate a variant of the original sequence on the combination of periodic Glycine features and the non-periodic Glycine feature, observing that both increase the probability of Glycine on the steered position. As with the original sequence, we again see that steering has no positive impact on the effect of the Non-periodic Glycine (rather it has a small, negative impact). 

\begin{figure}[htbp]
    \centering
    \includegraphics[width=0.8\linewidth]{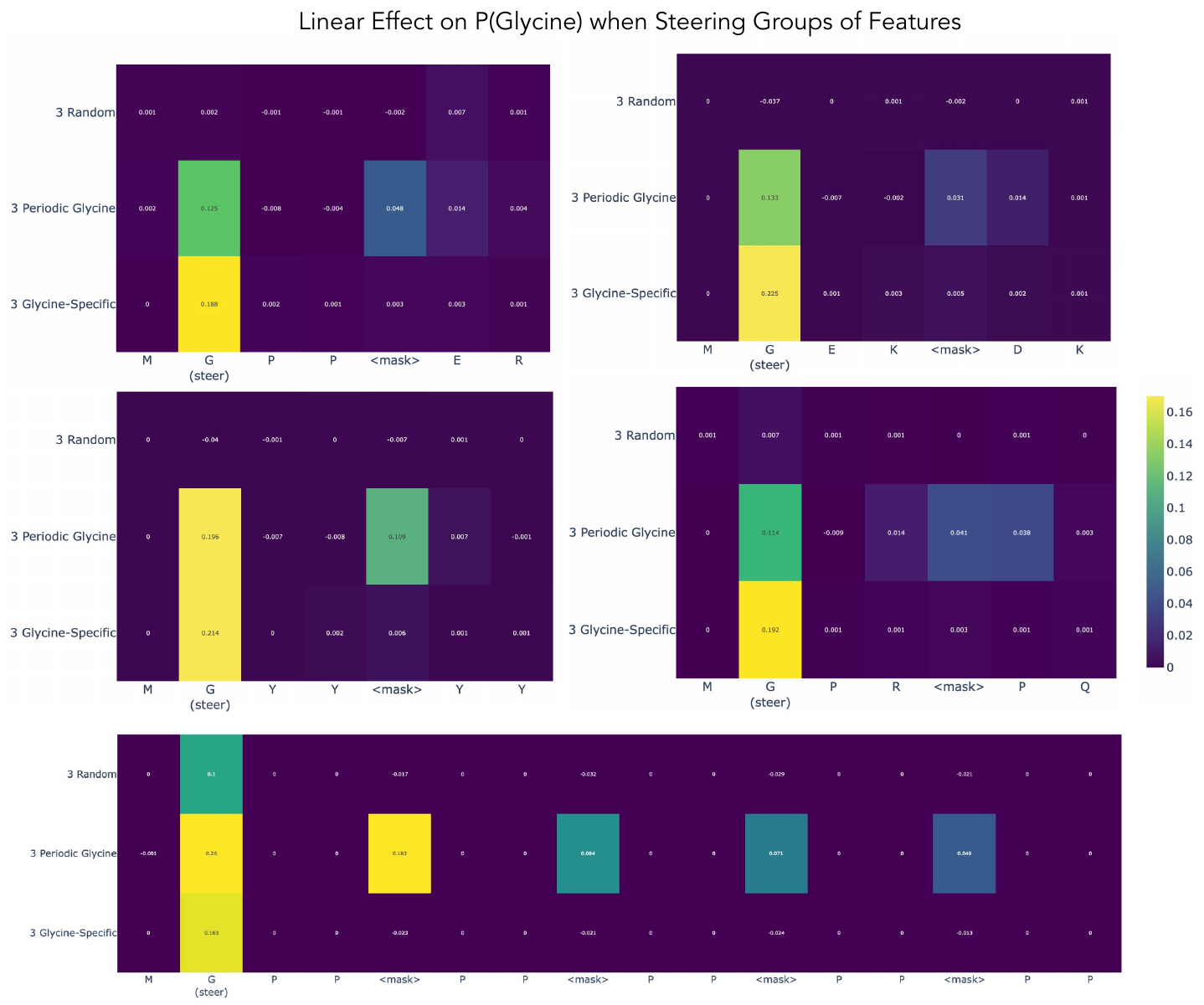}
    \caption{Measuring the linear effect on the predicted probability of Glycine.
    Evaluated p(Glycine) at each position in a sequence as groups of features were steered on first Glycine position at varying levels (0, 0.5,0.75, 1, 1.5, 1.75, 2). Calculated slope of p(Glycine) with respect to steering amount and visualize this for four sequences across each group of features. 3 Random features (f/0, f/1000, f/10000) were selected based on no activation-based information, 3 Periodic Glycine features (f/4616, f/4970, f/10003) selected based on maximum activation on periodic glycine patterns in collagen, and 3 Glycine-Specific features (f/6581, f/781, f/5389) selected to have the highest F1 scores to Glycine.
    }
    \label{fig:suppl-steer-2}
\end{figure}

Again, in Figure \ref{fig:suppl-steer-2}, we now test the same 3 periodic Glycine features (4616,4970, 10003), along with the 3 features with highest F1 to Glycine (6581, 781, 5489), and 3 randomly selected features (0, 1,000, 10,000). Here we calculate the slope of p(Glycine) across steering increasing [0, 0.5,0.75, 1, 1.5, 1.75, 2]. Alternate variants of the originally steered sequence maintain similar results and again, we see that even when applying 3 features at a time, the periodic Glycine features have a stronger effect on the masked position's Glycine probability than random or Glycine-specific features. This is also maintained in the longer sequence with multiple the masked periodic patterns being steered until this effect diminishes around the 5th masked position.

It should be noted with all of these experiments that we have only tested a few, somewhat constrained sequences so far, and more work needs to be done to evaluate the contexts in which steering this feature can or cannot work.

\bibliographystyle{unsrt}
\bibliography{references}

\end{document}